\documentclass[traditabstract]{aa} 
\usepackage{graphicx}
\usepackage{subfigure}
\usepackage{txfonts}
\usepackage{longtable}
\usepackage{multirow}
\usepackage{lscape}
\usepackage{pdflscape}
\usepackage{array}
\usepackage{threeparttable}
\usepackage{placeins}
\usepackage{natbib,twoopt}
\usepackage[breaklinks=true]{hyperref} 
\bibpunct{(}{)}{;}{a}{}{,}             
\makeatletter
\let\do@mlinenumbers\relax
\let\linenumbers\relax

\makeatother

\def\m2s2{\hbox{\,m$^{2}$\,s$^{-2}$}} 
\def\Msun{\hbox{$\mathrm{M}_{\odot}$}}		
\def\Me{\hbox{$\mathrm{M}_{\oplus}$}}			
\def\Re{\hbox{$\mathrm{R}_{\oplus}$}}			
\def\Mjup{\hbox{$\mathrm{M}_{\rm Jup}$}}		

\def\mp{M_{\rm p}}
\def\rp{R_{\rm p}}
\def\mstar{M_{\star}}

\newcounter{refsysno}

\newcounter{refsysbibno}

\newcounter{reftrno}

\newcounter{reftrbibno}

\begin{document}

\title{Cold Jupiters and small planets: Friends, foes, or indifferent?} 
\subtitle{A search for correlations with the largest exoplanet samples}
\titlerunning{On the relation between cold Jupiters and small planets}
\authorrunning{Bonomo et al.}

\author{A.~S.~Bonomo \inst{1}
\and L.~Naponiello \inst{1} 
\and E.~Pezzetta \inst{2}
\and A.~Sozzetti \inst{1}
\and D.~Gandolfi \inst{2}
\and R.~Wittenmyer\inst{3}
\and M.~Pinamonti\inst{1}
}

\institute{
INAF - Osservatorio Astrofisico di Torino, via Osservatorio 20, 10025 Pino Torinese, Italy  
\and Dipartimento di Fisica, Universit\`a degli Studi di Torino, via Pietro Giuria 1, 10125 Torino, Italy
\and University of Southern Queensland, Centre for Astrophysics, West Street, Toowoomba, QLD 4350, Australia
}

\date{Received 7 October 2024 / Accepted 20 May 2025}

\offprints{\email{aldo.bonomo@inaf.it}}

\abstract{Determining whether there is any correlation between the presence of short-period small planets (SPs) with $P\lesssim100$~d ($a \lesssim0.4$~AU) and $1 < \mp < 20\,\Me$ and that of outer cold Jupiters (CJs) with $a=1-10$~AU and $\mp=0.5-20\,\Mjup$ around solar-type stars may provide crucial constraints on models of formation and/or migration of SPs. However, discrepant results regarding the occurrence rates of CJs in SP systems have been reported in the literature, with some recent studies suggesting a strong SP-CJ correlation, but only at super-solar metallicities and/or masses of the host stars. Here, we homogeneously recomputed the occurrence rates of CJs at average, sub-solar ($\rm[Fe/H]<-0.1$), solar ($\rm-0.1\le[Fe/H]\le0.1$), and super-solar ($\rm[Fe/H]>0.1$) metallicities, as well as at average and subintervals of stellar mass, namely 0.6-0.8, 0.8-1.0, and 1.0-1.2$\,\Msun$, using: (i) a carefully selected sample of 217 SP systems detected through transits or radial velocity, and (ii) a large comparison sample of 1167 solar-type stars, regardless of the possible presence of SPs. We determine the integrated occurrence rate of CJs in SP systems to be $f_{\rm CJ|SP}=11.1^{+2.5}_{-1.8}\%$ at average metallicity $\overline{\rm [Fe/H]}=-0.011\pm0.005$ and mass $\overline{M_\star}=0.916\pm0.012\,\Msun$; this is consistent with the estimated frequencies of CJs in both the comparison sample ($f_{\rm CJ}=9.8^{+0.9}_{-0.8}\%$ at $\overline{\rm [Fe/H]}=-0.072\pm0.009$ and $\overline{M_\star}=0.994\pm0.004\,\Msun$)  and the HARPS-N survey of transiting SP systems. We find a possible correlation ($f_{\rm CJ|SP} > f_{\rm CJ}$) only at super-solar mass and metallicity, namely $1.0 \leq M_\star < 1.2\,\Msun$ and $\rm [Fe/H]>0.1$, although with statistical confidence of less than $3\sigma$.
To test some theoretical predictions, we also searched for possible SP-CJ relations as a function of SP and CJ multiplicity, as well as SP composition, albeit with the inevitably limited current sample, and we found none. We show that the architectures of SP systems are not indifferent to the presence of CJs, because the multiplicity of SPs strongly depends on the CJ eccentricity, as expected from planetary dynamics. 
A more comprehensive understanding of the relation between SPs and CJs requires larger samples of SP systems. The increasing number of well-characterized systems and the anticipated discoveries from both the Gaia and PLATO missions will enable a definitive assessment of the impact of CJs on the formation of SPs. 
}

\keywords{Planetary systems --
Planets and satellites: formation -- 
Stars: statistics --
Stars: solar-type --
Methods: statistical -- 
Techniques: radial velocities.}

\maketitle 

\section{Introduction} 
\label{introduction} 
Whether the presence of short-period small planets (SPs) with $P\lesssim100$~d (corresponding to a semimajor axis of $a \lesssim0.4$~AU) and $1 < \mp < 20\,\Me$ around solar-type stars correlates with that of cold Jupiters (CJs) -- gaseous giant planets with $a\simeq 1-10$\,AU and mass $0.5 \leq \mp \leq 20\,\Mjup$ (or $0.3 \leq \mp \leq 13\,\Mjup$, depending on the definition) -- is an intriguing question in exoplanetary science because it could potentially reveal crucial features of planet formation and evolution. However, there are strong discrepancies about this possible correlation in the literature from both theoretical and observational perspectives. 

Theoretically, an anticorrelation between SPs and CJs is expected if CJs either act as a dynamical barrier to the inward migration of icy SPs from beyond the water snowline at $\sim1-3$\,AU \citep{2015ApJ...800L..22I}, and/or if they considerably reduce the inward flux of pebbles to the inner regions of the protoplanetary disk, which is required to form dry SPs within the water snowline \citep{2019A&A...627A..83L}. From the Generation~3 Bern simulations of planet formation and evolution, \citet{2021A&A...656A..71S} instead find no or a weak SP-CJ correlation and predict a strong architecture-composition link, with more icy (rocky) SPs in the absence (presence) of one or more CJs, due to the same dynamical barrier effect of the CJ \citep{2015ApJ...800L..22I}. Furthermore, \citet{2021A&A...656A..71S} predict a scarcity of SPs at high stellar metallicity due to the disruption of inner planetary systems by more dynamically active CJs (see their Fig.~19). A recent study by \citet{2023A&A...674A.178B} shows that a correlation between the presence of SPs and that of CJs is possible, depending on the gas contraction rate for the formation of CJs: for slow gas contraction rates, the cores that form in the proximity of the water snowline are too small to effectively accrete large envelopes and can migrate inward, becoming SPs, while cores at greater distances can grow into cold gas giants. 

Observationally, occurrence rates of CJs in SP systems ($f_{\rm CJ|SP}$\footnote{$f_{\rm CJ|SP}$ is defined as the fraction of FGK dwarfs with at least one SP, which also host one or more CJs.}) as high as $\sim40\%$ reported by both \citet{ZhuWu2018} and \citet{bryan2019} disagree with a more recent estimate of $f_{\rm CJ|SP}=8.7^{+7.3}_{-2.7}\%$ for $\mp =0.5-20\,\Mjup$ (or $f_{\rm CJ|SP}=9.3^{+7.7}_{-2.9}\%$ for $\mp =0.3-13\,\Mjup$) by \citet{Bonomo2023} (hereafter B23). To reconcile these estimates, both \citet{2024RAA....24d5013Z} and \citet{2024ApJ...968L..25B} (hereafter BL24) explore the SP-CJ relation as a function of stellar metallicity and report a correlation at super-solar metallicity ($\rm[Fe/H]>0$). Specifically, BL24 analyzed a large sample of 184 SP systems detected by either transit or radial velocity (RV) surveys and compared their derived $f_{\rm CJ|SP}$ with the frequency of CJs ($f_{\rm CJ}$\footnote{$f_{\rm CJ}$ is defined as the fraction of FGK dwarfs hosting at least one CJ, regardless of the presence or absence of SPs.}) from the California Legacy Survey (CLS, \citealt{Rosenthal2021}). For this purpose, they considered the confirmed CJs and homogeneously computed the sensitivity to CJs for both their sample and the CLS survey. They find no correlation at $\rm [Fe/H] \leq 0$ 
and an SP-CJ correlation at $\rm[Fe/H]>0$ with a  $2.7\sigma$ significance level ($f_{\rm CJ|SP}=28.0^{+4.9}_{-4.6}\%$ vs $f_{\rm CJ}=14.3^{+2.0}_{-1.8}\%$). 
After submission of this paper, \citet{BryanLee2025} (hereafter BL25) report a positive correlation with a significance level greater than $2\sigma$ in metal-rich stars, with mass $M_\star>0.8\,\Msun$. 
BL25 interpret this as evidence for the critical role of stellar metallicity and/or mass (regarded as proxies for the disk solid and gas budget, respectively) in nucleating both inner low-mass and outer large-mass planets.

The studies by \citet{2024RAA....24d5013Z}, BL24, and BL25 were partly motivated by the observation that several stars in the \emph{Kepler} and K2 sample from B23 are metal-poor, which might bias the B23 estimate of $f_{\rm CJ|SP}$ toward lower values for the well-known relation between giant planets and stellar metallicity (e.g., \citealt{2005ApJ...622.1102F, Johnson2010, 2011A&A...533A.141S, 2013A&A...551A.112M}). However, the average metallicity of the B23 sample ($\overline{\rm [Fe/H]}=-0.066\pm0.014$) and that of the Anglo Australian Telescope (AAT) RV survey ($\overline{\rm [Fe/H]}=-0.045\pm0.018$) \citep{Wittenmyer2020}, which B23 used as a comparison sample, are consistent within $1\sigma$ and are both lower than the peak metallicity of approximately $-0.1$\,dex in a volume-limited sample of the solar neighborhood (e.g., \citealt{2011A&A...533A.141S}).

Based on the work of BL24, in the present study we recompute: (i) $f_{\rm CJ|SP}$ with a larger and more carefully selected sample of SP systems and (ii) $f_{\rm CJ}$, using three different blind RV campaigns, namely the AAT, CLS, and HARPS \citep{2011arXiv1109.2497M, Trifonov2020} surveys, and merging the stars of these three surveys into a single large sample. We then compare $f_{\rm CJ|SP}$ and $f_{\rm CJ}$ to search for possible SP-CJ correlations as a function of stellar metallicity, mass, or planet properties and draw our conclusions about their possible existence.

\section{Samples and radial velocity data}
\label{allsamples}

\subsection{Small planet transit and radial velocity sample} 
\label{small_planet_sample} 
The sample of SP systems to compute $f_{\rm CJ|SP}$ was built through queries to the NASA Exoplanet Archive\footnote{https://exoplanetarchive.ipac.caltech.edu/ .} on 2 December 2024 following the same criteria as BL24. The systems must have (1) at least one confirmed SP with $ 1 < \mp < 20\,\Me$\footnote{Unlike BL24, we did not consider the criterion $\rp \le 4\,\Re$ as this information is not available for the RV planets.} for transiting planets and $ 1 < \mp\sin{i} < 20\,\Me$ for planets discovered by RV surveys; (2) a solar-type (i.e., FGK dwarf or sub-giant) host star, thereby excluding M dwarfs with $\mstar \leq 0.6\,\Msun$; and (3) at least one publicly available RV dataset with more than 20 measurements and baseline $\Delta T\geq1$\,yr. 
However, unlike BL24, we initially discarded ``mixed systems'' containing both SP and short-period ($P < 100$~d) planets with $\mp \geq 20\,\Me$, such as the WASP-47 system, which also hosts a hot Jupiter (WASP-47b) between the two SPs WASP-47e and d (e.g., \citealt{Vanderburg2017, Nascimbeni2023}). It is unclear whether the theoretical predictions mentioned in Sect.~\ref{introduction} apply to such systems in which giant planets, especially the hot Jupiters, may have undergone a substantial migration process (e.g., \citealt{DawsonJohnson2018}). Nonetheless, we also computed the occurrence rates by taking the mixed systems into account and noted no significant differences (see Sect.~\ref{statistical_analyses}).
We removed a few outliers from the BL24 sample, such as TOI-1853b, which has $\mp=73 \pm 3>20\,\Me$ \citep{2023Natur.622..255N}, and Kepler-22b, whose orbital period is $P=290 > 100$\,d \citep{Borucki2012}. 

Our SP sample comprises 217 systems, of which 134 were detected by transit surveys and 83 by RVs. It is therefore considerably larger than that of BL24 (184 mixed systems) and contains 148 systems in common (see Table~1 in \href{http://doi.org/10.5281/zenodo.15855983}{Additional Tables}). Including the mixed systems leads to 242 systems, 151 and 91 of which were found with transits and RVs, respectively (Tables~1 and 2 in \href{http://doi.org/10.5281/zenodo.15855983}{Additional Tables}).
As in BL24, for each system we selected the available RV dataset with the best sensitivity for the detection of CJs (see Sect.~\ref{statistical_analyses}).

\subsection{Radial velocity comparison samples}
\label{comparison_samples}
As comparison samples to estimate $f_{\rm CJ}$, we considered the three largest RV surveys with high-precision RVs known to us, namely the AAT, CLS, and HARPS surveys, after applying the same criteria (2) and (3) as above for consistency. This yielded 196, 402, and 782 solar-type stars for the AAT, CLS, and HARPS campaigns, respectively. The number of solar-type stars from the CLS survey (402) is lower than that considered by BL24 (562), likely because BL24 applied the third selection criterion only to their SP sample and not to the CLS sample. Applying the third criterion (Sect.~\ref{small_planet_sample}) to the CLS survey reduced the number of solar-type stars by 160, while excluding only one system out of 52 with CJs, showing that a baseline $\Delta T \geq\,1$\,yr and more than 20 RV measurements for the best dataset of a given target are key to discovering CJs. The reduced number of CLS stars resulted in a slightly higher estimate of $f_{\rm CJ}$ (see Table~\ref{table_CLSsurvey_occurrence_rates}), which is more consistent with the $f_{\rm CJ}$ determinations from the other two surveys,  AAT and HARPS (see Table~\ref{table_RVsurveys_occurrence_rates}).

The mass, metallicity, and uncertainty on the metallicity of the selected solar-type stars from the three surveys were all taken from the Hypatia catalog\footnote{https://www.hypatiacatalog.com/ .} \citep{2014AJ....148...54H} and are listed in Table~3 of \href{http://doi.org/10.5281/zenodo.15855983}{Additional Tables}.

We also merged the three aforementioned RV surveys to create a large comparison sample with 1167 stars, 213 of which are common to two or all three surveys. In this way, we could compute a more precise and accurate $f_{\rm CJ}$ (see Tables~\ref{table_occurrence_rates_metstar} and \ref{table_RVsurveys_occurrence_rates}).

\subsection{Comparison of the small planet and radial velocity merged samples}
\label{comparison_SP_MegaRV_samples}

The distributions of the stellar metallicity and mass for both the SP and RV merged samples are compared in Fig.~\ref{figure_metmstardistr_SP_MegaRV}. 

The Kolmogorov-Smirnov (K-S) test between the two empirical cumulative distribution functions (eCDF) of metallicity yields a relatively low p-value of 0.021, as the SP sample contains a lower fraction of metal-poor stars at $\rm [Fe/H] \lesssim -0.2$ than the comparison sample (see Fig.~\ref{figure_metmstardistr_SP_MegaRV}, left panel).

The same K-S test for the stellar mass cumulative distributions returns a p-value of $3\mbox{\sc{e}-09}$, indicating that the two distributions are highly incompatible. This is due to an excess of lower-mass stars in the SP sample (see Fig.~\ref{figure_metmstardistr_SP_MegaRV}, right panel), given that SPs are more easily detected around smaller stars with both the transit and RV methods.

In Fig.~\ref{figure_metdistr_massbins_SP_MegaRV} we show the stellar metallicity distributions and eCDFs of both the SP and RV merged samples for the three intervals of stellar mass considered in Sect.~\ref{statistical_analyses}, namely $0.6-0.8$, $0.8-1.0$, and $1.0-1.2\,\Msun$. While the eCDFs of metallicity are compatible in the $0.6-0.8\,\Msun$ and $0.8-1.0\,\Msun$ intervals, with p-values of 99\% and 27\%, respectively, they differ considerably at super-solar mass (p-value~$< 1\%$) because of the considerably larger number of metal-poor stars in the RV merged sample (right panels of Fig.~\ref{figure_metdistr_massbins_SP_MegaRV}).

\section{Statistical analyses}
\label{statistical_analyses}
We determined $f_{\rm CJ|SP}$ from the SP sample (Sect.~\ref{small_planet_sample}) and $f_{\rm CJ}$ from both the individual RV surveys and the comparison sample (Sect.~\ref{comparison_samples}), first at the average metallicity of the different samples,
and then by considering three intervals in stellar metallicity, i.e., sub-solar ($\rm [Fe/H] < -0.1$), solar ($\rm -0.1 \le [Fe/H] \le 0.1$), and super-solar ($\rm [Fe/H] > 0.1$) metallicity. 
Similarly to BL24, we also split the samples into two broad intervals, namely $\rm [Fe/H] \le 0$ and $\rm [Fe/H] > 0$. 
However, given the typical uncertainties of $\sim0.04-0.07$~dex on $\rm [Fe/H]$, a star with $0.0 < \rm [Fe/H] < 0.1$ ($-0.1 < \rm [Fe/H] < 0.0$) should be better classified as solar in metallicity rather than as metal-rich (poor).

Given the different distributions of stellar mass in the SP and comparison samples (Sect.~\ref{comparison_SP_MegaRV_samples}), we further considered three different intervals in stellar mass, namely $ 0.6 \leq M_\star < 0.8\,\Msun$, $ 0.8 \leq M_\star < 1.0\,\Msun$, and $ 1.0 \leq M_\star < 1.2\,\Msun$\footnote{BL25 also considered similar intervals of stellar mass to investigate the SP-CJ relation as a function of stellar metallicity and mass.}. 
This is of interest because the frequency of giant planets also depends on stellar mass, though to a lesser extent than on metallicity (e.g., \citealt{Johnson2010, Fulton2021}).

For each sample and its relative subsamples in metallicity, we used binomial statistics:

\begin{equation}
b(d|N_{\star,\rm eff},f_{\rm CJ|SP})=\frac{N_{\star,\rm eff}!}{d!(N_{\star,\rm eff}-d)!}f_{\rm CJ|SP}^d(1-f_{\rm CJ|SP})^{N_{\star,\rm eff}-d},
\label{binom_equation}
\end{equation}

\noindent
where $d$ is the number of systems with at least one confirmed CJ, and $N_{\star,\rm  eff}$ is the ``effective'' number of stars, given by the product of the number of stars in the survey, $N_\star$, and the average survey sensitivity (or completeness) to CJs, $C$, i.e., $N_{\star,\rm  eff} = N_\star \cdot C$. 

The completeness of each individual star, $C_{\rm i}$, was computed in the same way as in B23 and BL24 (see their Sect. 3), 
that is, through injection and recovery experiments of CJ RV signals in a logarithmic grid of 30x30 $\Delta M_{\rm p}$-$\Delta a$ cells covering the ranges $0.3-20\,\Mjup$ and $1-10$\,AU, respectively. Specifically, for each cell, we simulated 200 RV signals of CJs at the epochs of the RV dataset by 
randomly varying: 
(i) $M_{\rm p}$ and $a$ within the cell bounds, 
(ii) $T_{\rm c}$ within the range of $P$ corresponding to $a$ and the stellar mass $M_\star$ 
(Tables~1-3 in \href{http://doi.org/10.5281/zenodo.15855983}{Additional Tables}) from Kepler's third law;  
(iii) $\cos{i}$ from 0 to 1, where $i$ is the orbital inclination; and 
iv) the argument of periastron $\omega$ from 0 to $2\pi$, while drawing the orbital eccentricity $e$ from a $\beta$ distribution \citep{2013MNRAS.434L..51K}. 
We then shifted every RV point at time $t$ following a Gaussian distribution with mean and standard deviation equal to the RV value and its uncertainty, respectively. We estimated $C_{\rm i}$ as the recovery rate of the simulated signals (Keplerian orbits or linear and quadratic trends if $\Delta T$ was considerably shorter than the simulated CJ orbital period) across the whole grid using the $\Delta BIC$ criterion (see B23 and BL24). The average completeness $C$ of each sample (or subsample) was then obtained by the mean of the completenesses of the stars in that sample: $C=(1/N_{\star})\sum_{i=1}^{N_\star}{C_{\rm i}}$. 
The average completenesses of all the samples described in Sect.~\ref{allsamples} are displayed in Fig.~\ref{fig:all_completenesses}. We note a very good agreement when comparing our average completenesses of the SP RV and transit samples (upper panels in Fig.~\ref{fig:all_completenesses}) with those shown in Fig.~A.1 of BL24 (for $a=1-10$~AU).

We considered two ranges of CJ mass\footnote{Note that only minimum masses, $\mp\sin{i}$, are known for the majority of CJs as their orbital inclination $i$ cannot be determined from RVs alone.}, 
namely $\mp=0.5-20\,\Mjup$ and $\mp=0.3-13\,\Mjup$ (for comparison with B23), but we mainly rely on the former because the latter suffers from incompleteness issues between 0.3 and 0.5$\,\Mjup$ (Fig.~\ref{fig:all_completenesses}). For each CJ mass range and the aforementioned bins of stellar metallicity and/or mass, we provide the number of solar-type stars ($N_\star$), the number of stars with confirmed CJs ($d$), the average completenesses ($C$), and CJ occurrence rates from (i) the SP and the large comparison samples in Tables~\ref{table_occurrence_rates_metstar} and \ref{table_occurrence_rates_mstar_metstar}, (ii) the SP transit, RV, and transit+RV samples as well as the SP and mixed transit, RV, and transit+RV samples in Table~\ref{table_SP_Mixed_occurrence_rates}, and (iii) the AAT, CLS, and HARPS surveys in Table~\ref{table_RVsurveys_occurrence_rates}.
Tables~4, 5, and 6 in \href{http://doi.org/10.5281/zenodo.15855983}{Additional Tables} list the known CJs and their properties in the SP, mixed, and comparison samples. Figure~\ref{fig:SP_CJ} shows the architectures of systems in the SP sample: on the left, the 27  systems with CJs  of mass $\mp=0.3-13\,\Mjup$, and on the right, the 23 systems with CJs of mass $0.5-20\,\Mjup$.

\section{Results} 
\label{results} 
\subsection{Frequency of CJs at average stellar metallicity and mass}
At the average stellar metallicity ($\overline{\rm [Fe/H]}=-0.011\pm0.005$) and mass ($\overline{M_\star}=0.916\pm0.012\,\Msun$) of the SP (transit+RV) stars, we determine $f_{\rm CJ|SP}=11.1^{+2.5}_{-1.8}\%$ for the CJ mass range $\mp=0.5-20\,\Mjup$. Including the 25 mixed systems yields $f_{\rm CJ|SP+Mixed}=12.5^{+2.5}_{-1.8}\%$.
Returning to the original sample of BL24, which has $\overline{\rm [Fe/H]}=-0.001\pm0.005$ and $\overline{M_\star}=0.923\pm0.014\,\Msun$ and includes the mixed systems, and accounting for its average completeness\footnote{This was extracted from Fig.~A.1 in BL24.}, we find a fully consistent result: $15.2^{+3.1}_{-2.3}\%$. Both occurrence rates are lower than previous claims of $f_{\rm CJ|SP}\sim 40\%$ (Sect.~\ref{introduction}). They agree within $1\sigma$ with the estimate of $f_{\rm CJ|SP}=8.7^{+7.3}_{-2.7}\%$ by B23, whose smaller sample is on average slightly more metal-poor by $\sim 0.06$~dex. 

For the comparison sample, at its average metallicity ($\overline{\rm [Fe/H]}=-0.072\pm0.009$) and mass ($\overline{M_\star}=0.994\pm0.004\,\Msun$), we find $f_{\rm CJ}=9.8^{+0.9}_{-0.8}\%$, which is fully compatible with $f_{\rm CJ|SP}$\footnote{The possible, though not significant, hint for an SP-CJ anticorrelation reported by B23 can be ruled out as our $f_{\rm CJ}$ recomputed from the AAT survey is approximately half that derived from \citet{Wittenmyer2020}.}.

From our occurrence rates $f_{\rm CJ|SP}$ and $f_{\rm CJ}$, and the frequency of SPs $f_{\rm SP}=28.1^{+6.6}_{-5.1}\%$ from \citet{Rosenthal2022} (see also \citealt{Zhu2018}), 
we can derive the conditional probability that a solar-type star with a CJ also hosts a SP \citep{ZhuWu2018}, namely $f_{\rm SP|CJ}=f_{\rm SP} \times f_{\rm CJ|SP}~/~f_{\rm CJ} = 32 \pm 11\%$. 
This agrees within $1\sigma$ with the $f_{\rm SP|CJ}$ computed from a sample of 115 solar-type stars known to host a CJ (Ruggieri et al., in prep.), and is substantially lower than the estimate of $f_{\rm SP|CJ} \sim 100\%$ in \citet{bryan2019}.

\subsection{Frequency of CJs as a function of stellar metallicity}
\label{frequency_CJs_metstar}
\begin{figure}[t!]
\centering
\includegraphics[width=1.0\columnwidth]{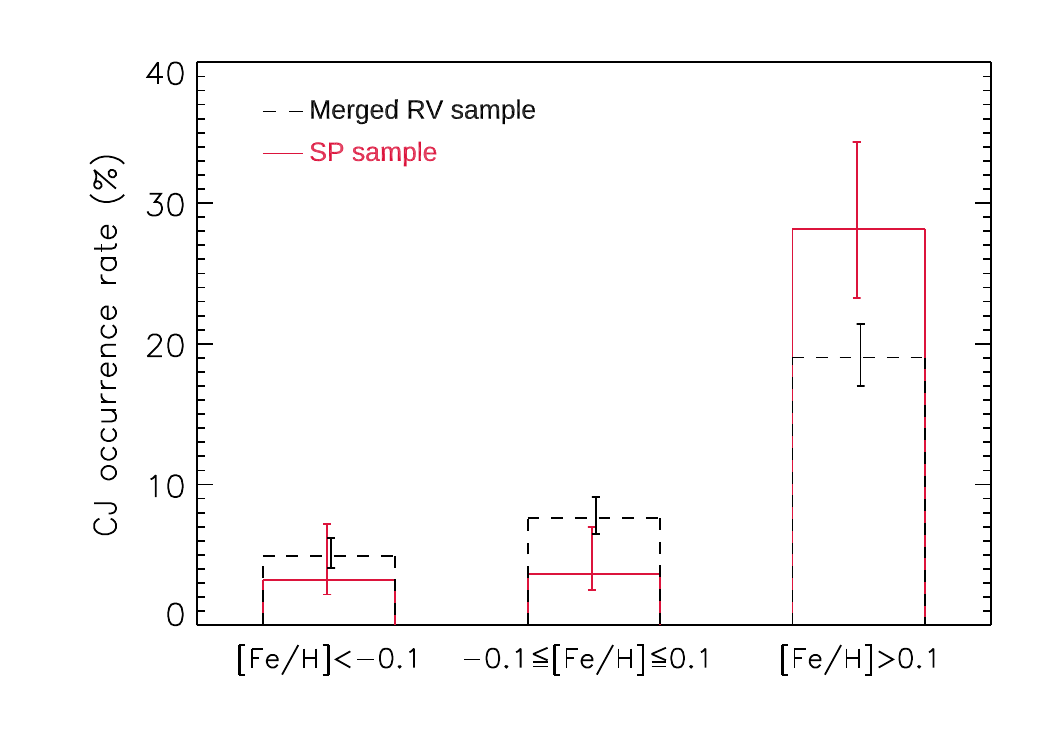} \\
\vspace{-0.5 cm}
\caption{Occurrence rates of CJs with $\mp = 0.5-20\,\Mjup$ in the SP sample (red solid line) and the merged RV comparison sample (black dashed line) at sub-solar ($\rm[Fe/H]<-0.1$), solar ($\rm-0.1\le[Fe/H]\le0.1$), and super-solar ($\rm[Fe/H]>0.1$) metallicity. See Table~\ref{table_occurrence_rates_metstar}. }
\label{figure_occurrence_rates_metstar}
\end{figure}

We report $f_{\rm CJ|SP}$ and $f_{\rm CJ}$ for the three subintervals of stellar metallicity (Sect.~\ref{statistical_analyses}) in Table~\ref{table_occurrence_rates_metstar}and display them in Fig.~\ref{figure_occurrence_rates_metstar} (see also Fig.~\ref{figure_occurrence_rates_metstar_03_13_Mjup} for $\mp=0.3-13\,\Mjup$). For the same metallicity bins, we also provide the estimates of $f_{\rm CJ}$ from the AAT, CLS, and HARPS surveys separately in Table~\ref{table_RVsurveys_occurrence_rates}; these are fully consistent with each other (see  Fig.~\ref{figure_occurrence_rates_RVsurveys}).

By comparing $f_{\rm CJ|SP}$ and $f_{\rm CJ}$ in Table~\ref{table_occurrence_rates_metstar}, we find that (i) at sub-solar and solar metallicities, they are fully compatible within $1\sigma$, and (ii) at super-solar metallicity, the value of $f_{\rm CJ|SP}$ is higher than $f_{\rm CJ}$ by approximately $3/2$ ($\sim 30\%$ vs $\sim 20\%$; Fig.~\ref{figure_occurrence_rates_metstar}), but the significance of this difference is low, at $1.7\sigma$ (compared to $2.7\sigma$ in BL24 for $\rm [Fe/H] > 0$).

\begin{figure*}[t!]
\centering
\includegraphics[width=0.725\columnwidth]{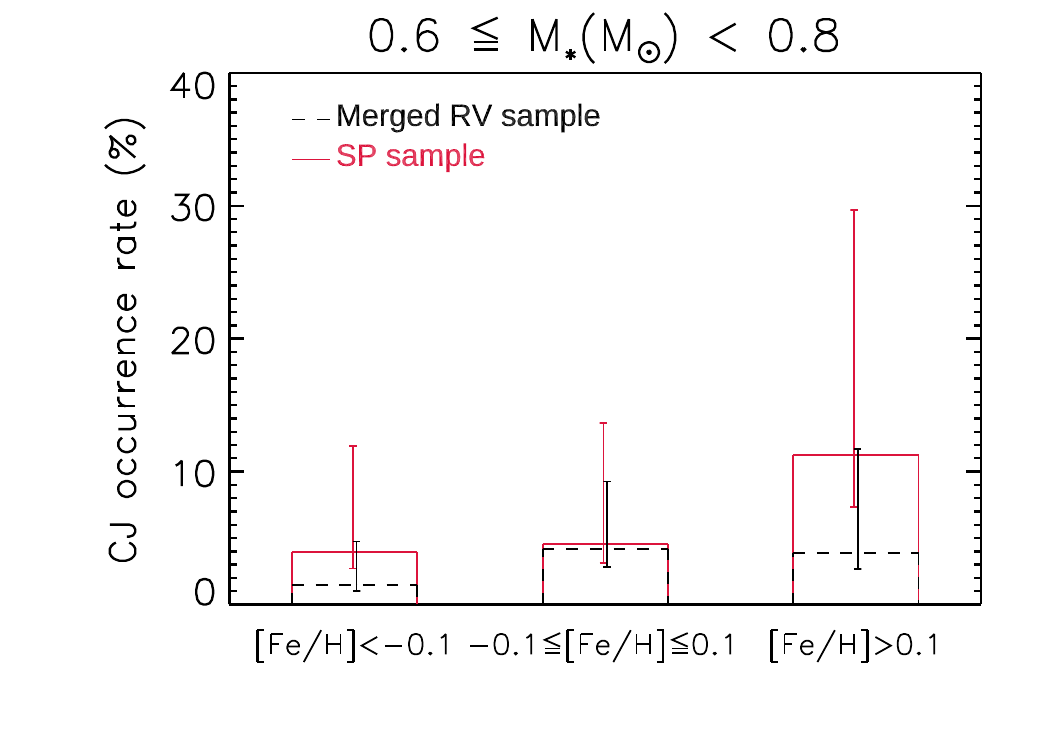}
\hspace{-0.75cm}
\includegraphics[width=0.725\columnwidth]{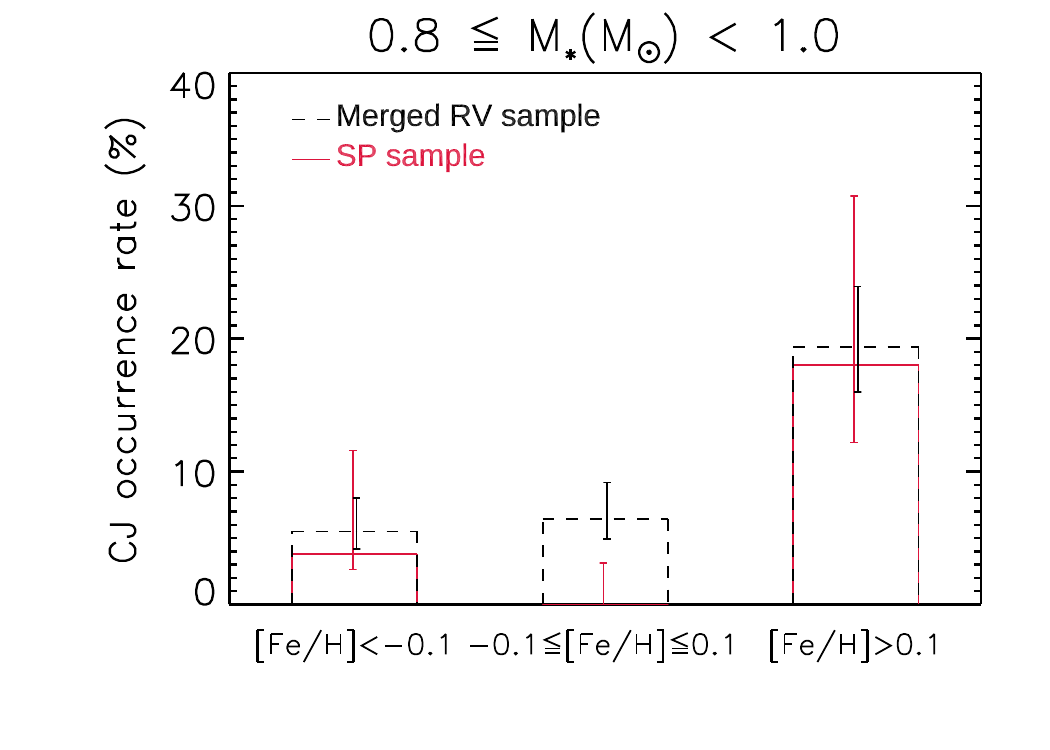}
\hspace{-0.75cm}
\includegraphics[width=0.725\columnwidth]{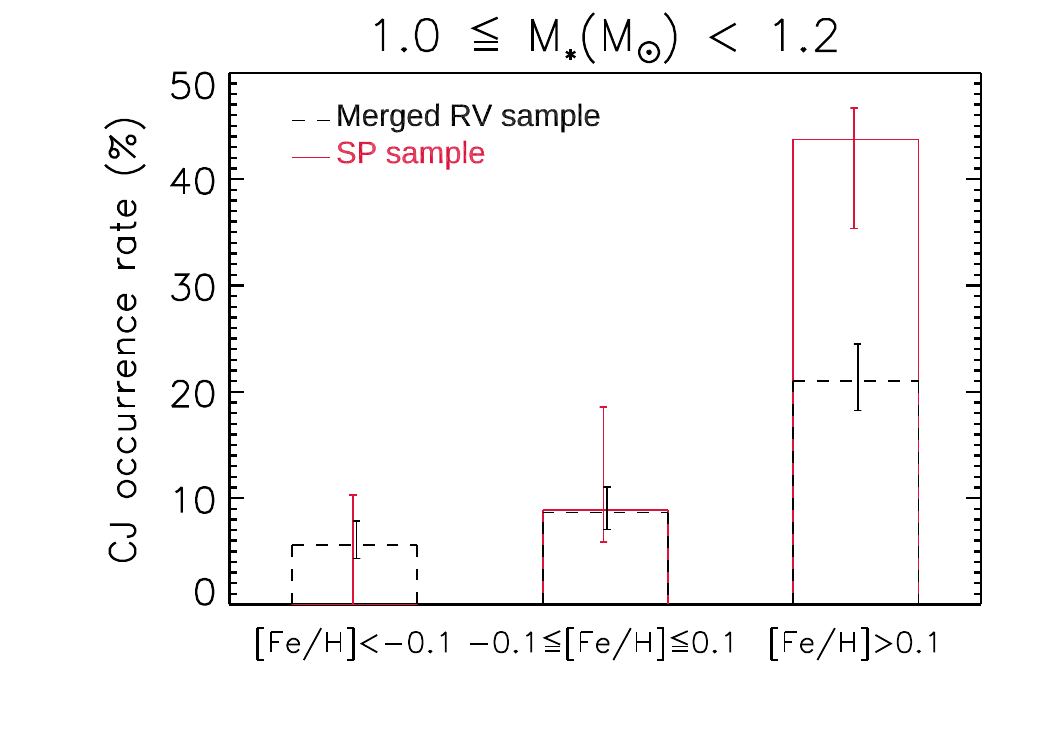} \\
\vspace{-0.5 cm}
\caption{Same as Fig.~\ref{figure_occurrence_rates_metstar} for three different bins of stellar mass: $0.6-0.8\,\Msun$ (left), $0.8- 1.0\,\Msun$ (center), and $1.0-1.2\,\Msun$ (right). See Table~\ref{table_occurrence_rates_mstar_metstar}.}
\label{figure_occurrence_rates_mstar_metstar}
\end{figure*}

\subsection{Frequency of CJs as a function of stellar metallicity and mass}
\label{frequency_CJs_metstar_mstar}
We report $f_{\rm CJ|SP}$ and $f_{\rm CJ}$ for the same bins of stellar metallicity as above and for the three ranges of stellar mass discussed in Sect.~\ref{statistical_analyses}, in Table~\ref{table_occurrence_rates_mstar_metstar}. We show these values in Fig.~\ref{figure_occurrence_rates_mstar_metstar}  (see also Fig.~\ref{figure_occurrence_rates_mstar_metstar_03_13_Mjup} for $\mp=0.3-13\,\Mjup$).

We find an excess of CJs in SP systems at $2.5\sigma$ significance level only for super-solar mass and metallicity, namely $M_\star \geq 1.0\,\Msun$ and $\rm [Fe/H] > 0$ 
(see the right panel in Fig.~\ref{figure_occurrence_rates_mstar_metstar}, and Table~\ref{table_occurrence_rates_mstar_metstar}), in agreement with the recent findings of BL25. This is the origin of the higher $f_{\rm CJ|SP}$ compared to $f_{\rm CJ}$ at super-solar metallicity and integrated stellar masses (Sect.~\ref{frequency_CJs_metstar} and Fig.~\ref{figure_occurrence_rates_metstar}).

\subsection{SP-CJ relations as a function of SP and CJ multiplicity}

We explored possible SP-CJ trends as a function of SP multiplicity. For this analysis, we adopted the $0.3-13\,\Mjup$ mass range (instead of $0.5-20\,\Mjup$), given the higher number of systems with both CJs and SPs (27 compared to 23).  If the presence of CJs hinders the formation or migration of SPs, one would expect a lower degree of SP multiplicity in the presence of CJs. For example, low-mass ``jumpers'' overtaking the CJ towards inner orbits are more likely to form single SP systems (see Fig.~3 in \citealt{2015ApJ...800L..22I}). By comparing the frequency of multiple systems with a number of SPs $n_{\rm p}\geq2$ in the SP subsamples with and without CJs, we find $f_{n_{\rm p}\geq2,\rm w/\,CJ}=37^{+10}_{-8}\%$ (10 out of 27 systems) and $f_{n_{\rm p}\geq2,\rm w/o\,CJ}=43^{+4}_{-3}\%$ (81 out of 190 systems), respectively. These rates are statistically consistent within uncertainties. 

We also searched for possible (anti)correlations as a function of CJ multiplicity to investigate whether two or more CJs may hinder SP formation more than a single CJ system (Sect.~\ref{introduction}). To this end, we compared the frequency of single SPs in systems with one CJ to that in systems with two (or more) CJs. The latter should be higher if two or more CJs have a more deleterious effect than a single CJ on the formation of multiple SP systems.
We find no evidence for this: the rates of single SP systems with one CJ and two CJs are compatible, namely $63^{+9}_{-12}\%$ (12 out of 19 systems) and $62^{+13}_{-18}\%$ (5 out of 8 systems), respectively (see Fig.~\ref{fig:SP_CJ}).

\subsection{SP-CJ relation as a function of SP composition} 
\label{freq_SP_composition}

We investigated the possible presence of an architecture-composition link by determining the frequency of rocky SPs\footnote{We define a rocky planet as one located between the 100\% iron and 100\% silicate compositions in the mass-radius diagram of small planets (e.g., \citealt{ZengSasselov2013}).} in CJ systems. This analysis tests whether the formation of rocky planets is favored in the presence of CJs if CJs block the migration of icy sub-Neptunes from beyond the water snowline \citep{ 2021A&A...656A..71S}. We derive $f_{\rm rocky,w/\,CJ}=33^{+13}_{-9}\%$ from five transiting systems with rocky planets (HD219134, Kepler-139, Kepler-407, K2-312, and K2-364) out of fifteen in total (including three RV systems that were later found transiting). This indicates that systems with both rocky planets and CJs do not represent the majority of the SP-CJ transiting systems.

\subsection{SP multiplicity as a function of CJ eccentricity} 
\begin{figure}[h]
\centering
\hspace{-0.5 cm} \includegraphics[width=0.51\textwidth]{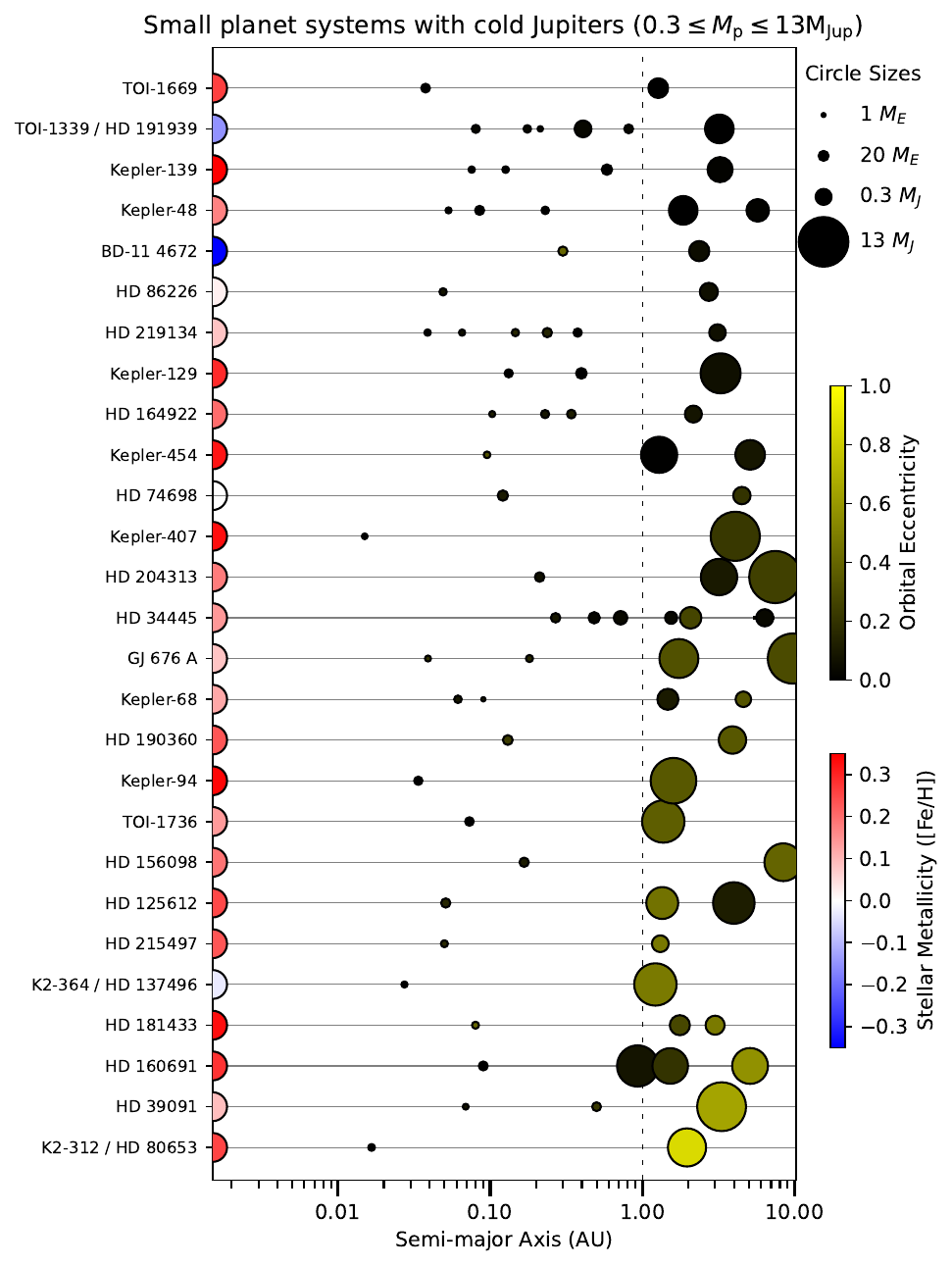}
\caption{Architectures of 27 systems with both SPs and CJs in the mass range $0.3\leq{\rm M_p}\leq13\,\Mjup$, ordered according to increasing orbital eccentricity of the CJs (from top to bottom; for multiple CJs, the highest eccentricity is considered). The planet circles have sizes proportional to their mass (see legend in the top right corner), and colors indicating orbital eccentricity (more eccentric planets are shown in yellow). Host stars are color coded as a function of stellar metallicity (metallicity increases from blue to red). 
All systems with three or more inner SPs are accompanied by CJs with low eccentricities.}
\label{fig:SP_CJ_ecc}
\end{figure}

When ordering the SP systems with CJs  as a function of the CJ eccentricity (Fig.~\ref{fig:SP_CJ_ecc}), only single or, more rarely, two SPs are found with eccentric ($e\gtrsim0.4$) CJs, the most extreme case being K2-312/HD\,80653 (B23). In contrast, higher multiplicity (up to $n_{\rm p}=4$ for HD\,219134) SP systems  are found in the company of CJs in quasi-circular orbits. This is consistent with the expectation of stronger dynamical perturbations at inner orbital separations by eccentric CJs (e.g., \citealt{2017AJ....153..210H}).

\section{Discussion and conclusions} 
\subsection{Cold Jupiters and small planets: friends, foes, or indifferent?} 
In this work, we first investigated the possible SP-CJ relation as a function of stellar metallicity and mass. This did not reveal any correlation or anticorrelation (i.e., SPs and CJs are neither friends nor foes) in eight of the nine mass and metallicity bins, that is, at sub-solar mass for any metallicity and at super-solar mass for solar and sub-solar metallicity. We instead find an enhancement of CJs in SP systems (i.e., SPs and CJs could be friends) at super-solar mass and metallicity, though at less than $3\sigma$, in general agreement with BL25.

We do not find any SP-CJ trends with planet multiplicity and composition, which further challenge the anticorrelation models predicting a dearth of CJs in SP systems (Sect.~\ref{introduction}). Nonetheless, it should be noted that (i) the limited subsamples of SP systems with CJs (27 systems) inevitably result in large uncertainties in CJ occurrence rates, making it difficult to detect possible weak correlations or anticorrelations with planet properties; and (ii) the SP multiplicity considered in this work is based on the detected SPs and may be different from the true multiplicity, especially if non-transiting SPs with relatively long periods ($P\sim25-100$~d) and/or low mass were not detected by RV follow-up.

Our investigation of the architecture-composition relation in Sect.~\ref{freq_SP_composition} must be considered with caution because it only concerned rocky planets. In principle, it may also apply to rocky planets with tenuous H/He atmospheres, if some available gas was accreted by the super-Earths at the time of their formation, and was not subsequently stripped away by post-formation processes such as atmospheric photoevaporation (e.g., \citealt{2014ApJ...792....1L}) or core-powered mass loss (e.g., \citealt{2018MNRAS.476..759G}). 
However, the well-known degeneracy in SP composition from the measurement of the bulk density alone does not allow us to distinguish whether or which non-rocky planets in the ten transiting systems with CJs of our SP sample are actually dry (rocky worlds shrouded by thin H/He atmospheres, which formed within the water snowline) or icy (water worlds with possible H/He envelopes, formed beyond the water snowline and then migrated inwards). This makes the investigation of the architecture-composition link challenging.

We show that higher multiplicity SP systems are found in systems with CJs in low-eccentricity orbits, being immune to dynamical instabilities generated by highly eccentric CJs. In this sense, SPs are not indifferent to CJs, even in the majority of cases where $f_{\rm CJ | SP} \sim f_{\rm CJ}$.

\subsection{Caveats of this work, BL24, and BL25}
The occurrence rates derived in this work, BL24, and BL25 may be slightly underestimated due to two primary factors. However, these factors should affect both the SP and comparison samples approximately equally, making the search for correlations, in principle, reliable. Firstly, the completeness values computed in our work and in BL24 are optimistic because they only account for the formal RV uncertainties and not for uncorrelated jitter (as addressed comprehensively by B23), which in some cases may be considerably higher than the former\footnote{Nevertheless, proper Markov chain Monte Carlo analyses of the RV datasets of $\sim 1350$ systems considered here, similarly to B23, would require unrealistic computation times.}. 
Secondly, a small number of linear or quadratic RV trends, which might later result in CJ signals, were not taken into account in the computation of the occurrence rates, which is based only on confirmed CJs. Nevertheless, among the targets in the SP sample with relatively short temporal baselines, that is $1 \leq \Delta T \leq 4$\,yr, we identified only three targets with RV long-term trends still compatible with CJs (namely, TOI-1279, TOI-1742, and TOI-1778). Moreover, such trends are very often induced by brown dwarfs or low-mass stellar companions, or they correlate with activity indicators (see B23).

Conversely, occurrence rates may decrease if the minimum masses ($\mp\sin{i}$) of some of the considered CJs turn out to be true masses ($\mp$) above the 20 (or 13) $\Mjup$ upper limit, corresponding to brown dwarfs or low-mass stellar objects, due to low orbital inclinations (e.g., \citealt{Xiao2023}).

\subsection{Perspectives}
As more and more SP systems are characterized with high-precision RV follow-up and monitored over extended periods, more precise and accurate determinations of $f_{\rm CJ|SP}$ can be derived, allowing further investigation of the SP-CJ relation, even in the lower CJ mass range $0.3-13\,\Mjup$. We plan to update our results over time by incorporating the increasing number of well-characterized SP systems. 

Crucial progress in this regard will come from the Gaia (e.g., \citealt{2024CRPhy..24S.152S}) and PLATO \citep{2014ExA....38..249R, 2025ExA....59...26R} space missions. The Gaia mission, particularly with Data Release 4 anticipated in 2026, is expected to (i) provide crucial information on the presence of CJs (e.g., \citealt{2023A&A...674A..10H}), especially around relatively bright stars; (ii) provide CJ true masses (note that about $30\%$ of the CJs in the SP sample already have measured $\mp$ owing to the combination of RVs and Hipparcos-Gaia proper motion anomalies; see, e.g., \citealt{2022A&A...657A...7K, Errico2022}; and Table~4 in \href{http://doi.org/10.5281/zenodo.15855983}{Additional Tables}); and (iii) help us clarify the origin of certain RV long-term trends (e.g., \citealt{2021ApJ...922L..43B}). PLATO is expected to detect $\gtrsim 8,000$ SP systems orbiting bright solar-type stars with $V \leq 11$\footnote{This can be estimated from the planet detection rate in \citealt{2023A&A...677A.133M} after a 2+2 yr mission and the \emph{Kepler} transit multiplicity distribution in \citealt{2021ARA&A..59..291Z}.}, which will be well suited to high-precision RV follow-up. Moreover, PLATO will reveal new SPs in known TESS systems, enabling more in-depth studies of $f_{\rm CJ|SP}$ as a function of SP multiplicity.

\section{Data availability}
Tables listing the stars and their properties, as well as the cold Jupiters and their parameters, in the small planet, mixed, and 
comparison samples are available in \href{http://doi.org/10.5281/zenodo.15855983}{Additional Tables}.

\begin{acknowledgements} 
We are grateful to the anonymous referee for her/his valuable comments, which allowed us to considerably improve this work. We are thankful to Profs. A. Morbidelli and B. Bitsch for useful discussions about theoretical predictions on SP-CJ relations. 
We also thank Dr.~K.~Biazzo for her suggestion to use the Hypatia catalog. 
We acknowledge financial contribution from the INAF Large Grant 2023 ``EXODEMO'' and from the European Union - Next Generation EU RRF M4C2 1.1 PRIN MUR 2022 project 20229R43BH. D.G. gratefully acknowledges the financial support from the grant for
internationalization (GAND\_GFI\_23\_01) granted by the University of Turin
(Italy).
This research has made use of the NASA Exoplanet Archive, which is operated by the California Institute of Technology, under contract with the National Aeronautics and Space Administration under the Exoplanet Exploration Program. 
The research shown here acknowledges use of the Hypatia Catalog Database, an online compilation of stellar abundance data, which was supported by NASA's Nexus for Exoplanet System Science (NExSS) research coordination network and the Vanderbilt Initiative in Data-Intensive Astrophysics (VIDA).
\end{acknowledgements}

\bibliographystyle{aa} 
\bibliography{aa52523-24} 

\begin{appendix}

\twocolumn


\onecolumn

~\\

\section{Comparison of the stellar properties of the small planet and merged radial velocity samples}

\begin{figure}[h!]
\centering
\includegraphics[width=0.475\textwidth]{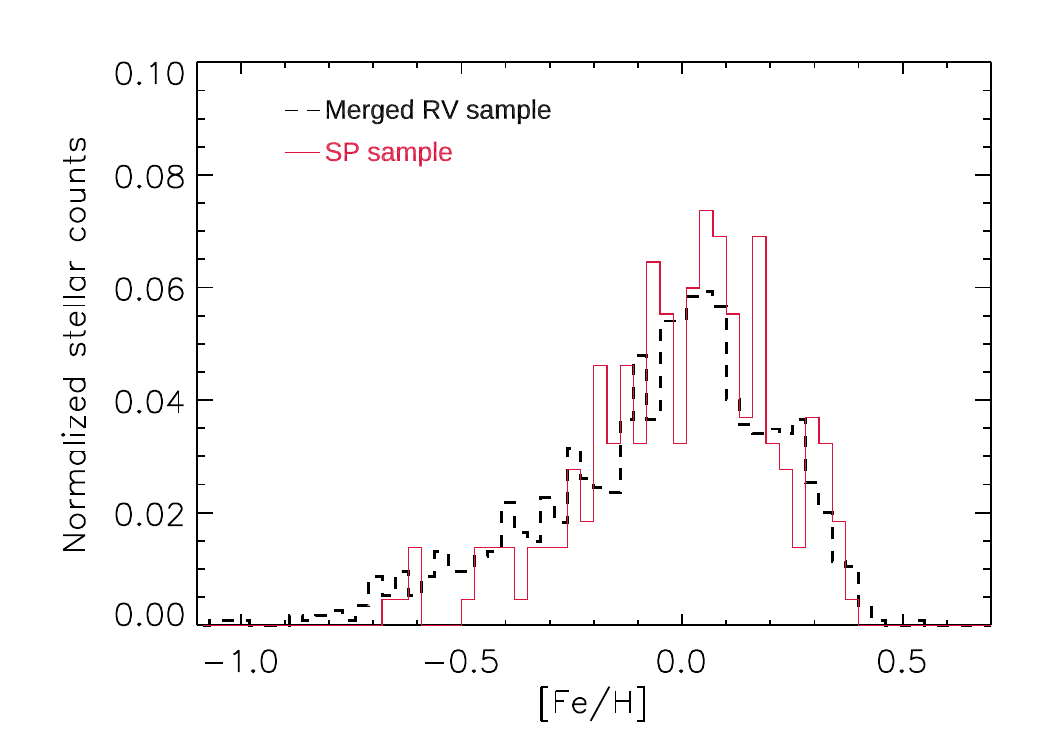}
\hspace{0.45 cm}
\includegraphics[width=0.475\textwidth]{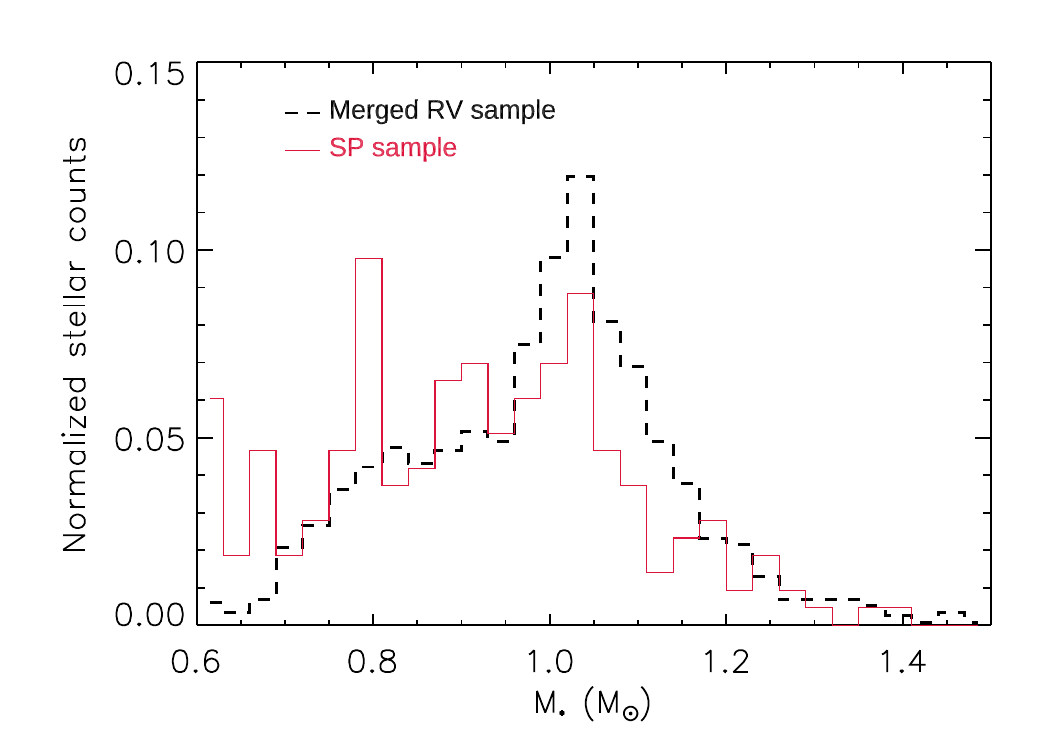}
\caption{Distributions of stellar metallicity (left panel) and stellar mass (right panel) in the SP sample (red solid line) and merged (AAT+CLS+HARPS) RV comparison sample (black dashed line).
Note that the relative stellar counts of the SP and RV merged samples cannot be directly compared as they were normalized from different distributions.}
\label{figure_metmstardistr_SP_MegaRV}
\end{figure}

~\\

~\\

\begin{figure}[h!]
\centering
\includegraphics[width=0.352\columnwidth]{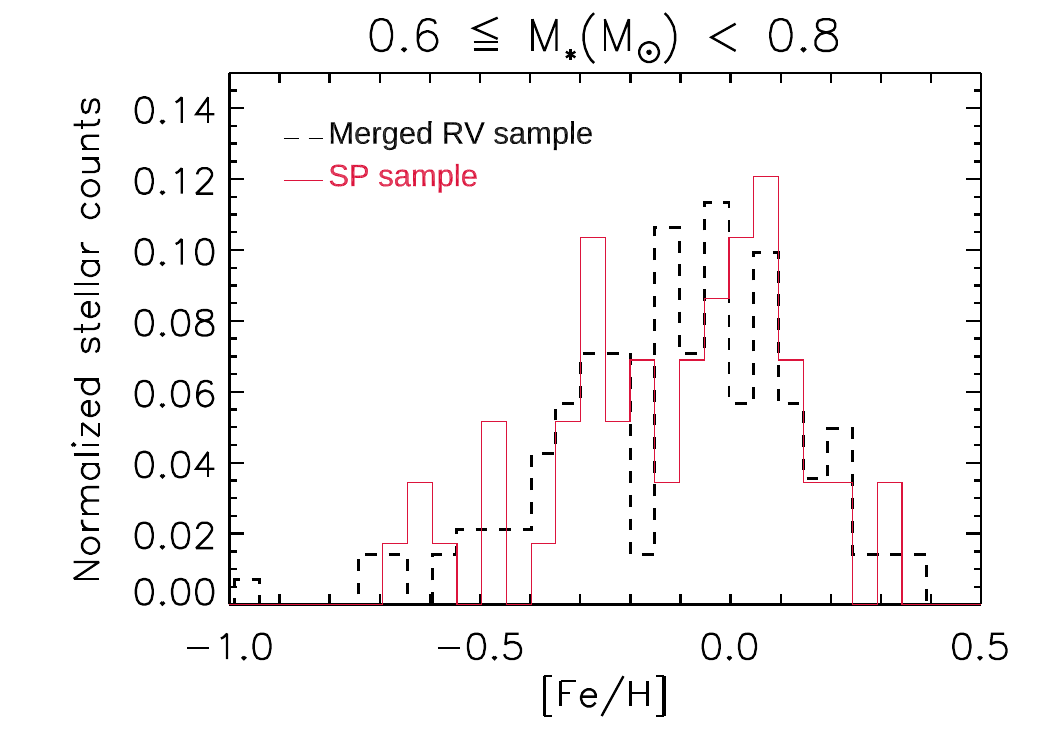}
\hspace{-0.75cm}
\includegraphics[width=0.352\columnwidth]{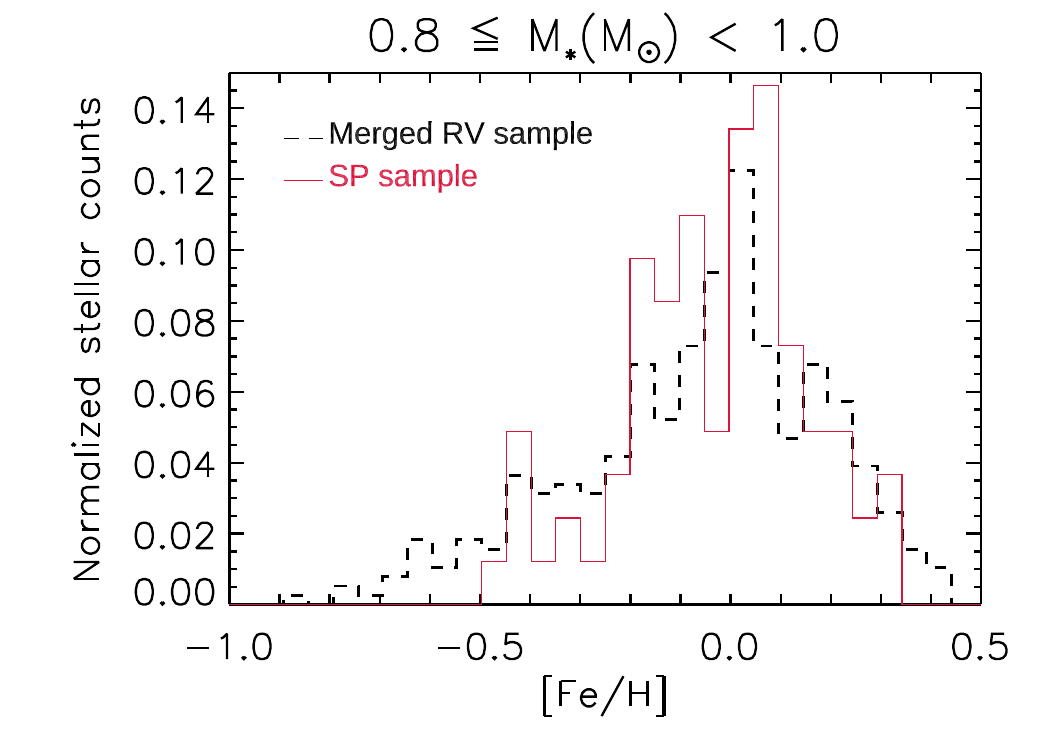}
\hspace{-0.75cm}
\includegraphics[width=0.352\columnwidth]{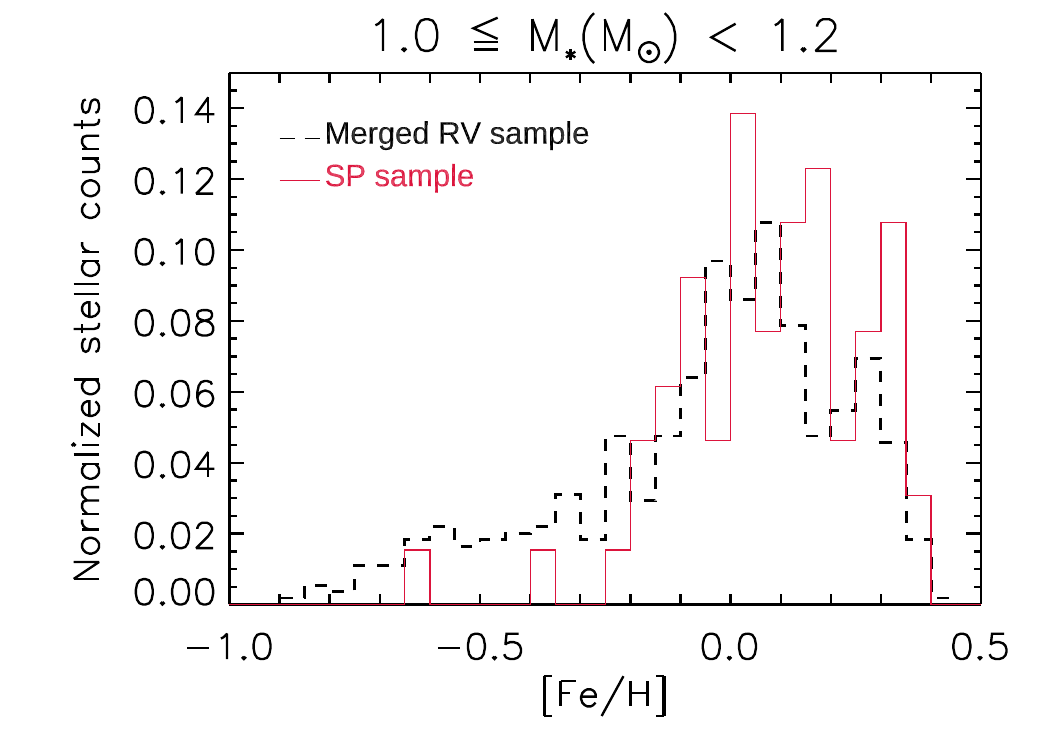} \\
\includegraphics[width=0.352\columnwidth]
{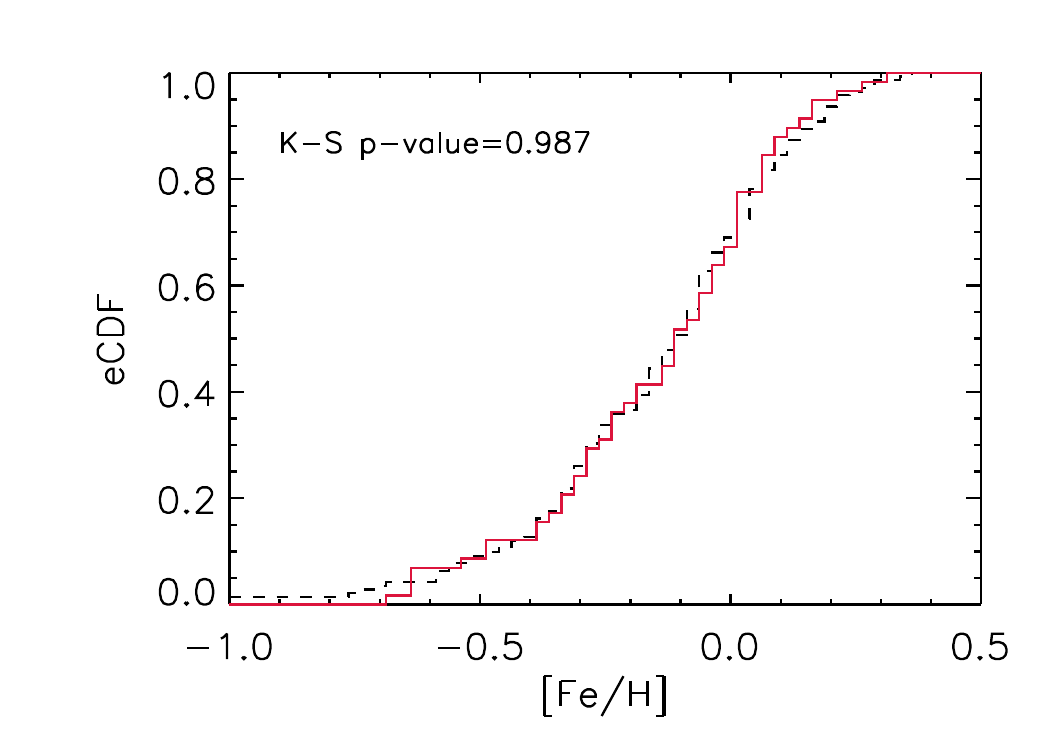}
\hspace{-0.75cm}
\includegraphics[width=0.352\columnwidth]{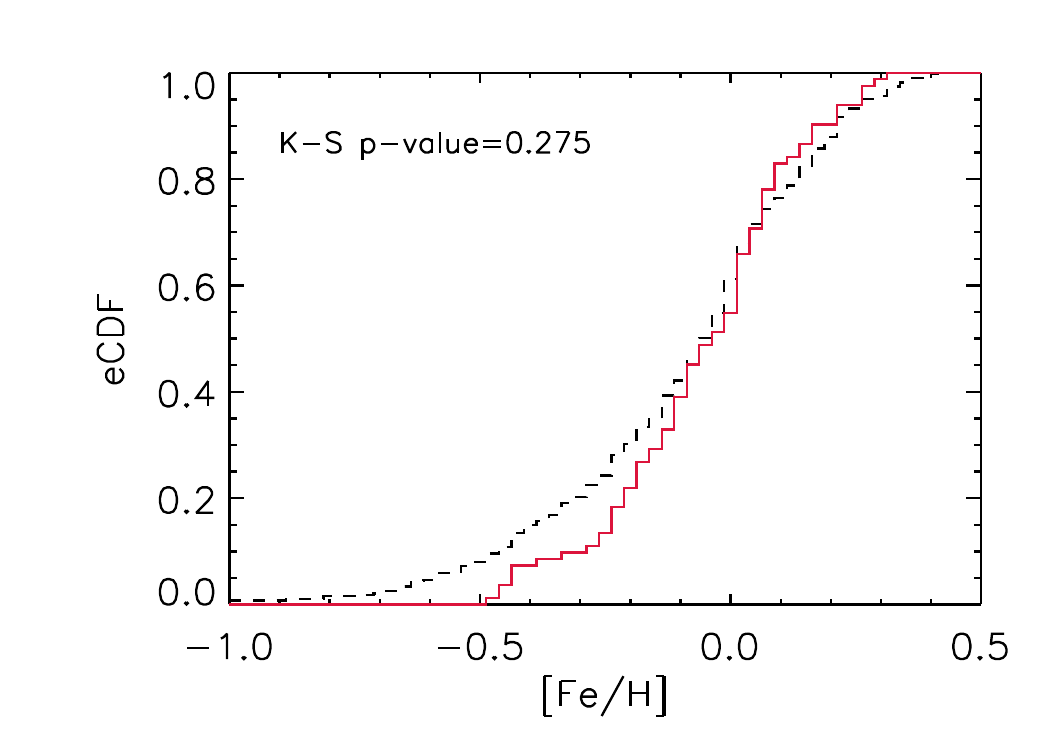}
\hspace{-0.75cm}
\includegraphics[width=0.352\columnwidth]{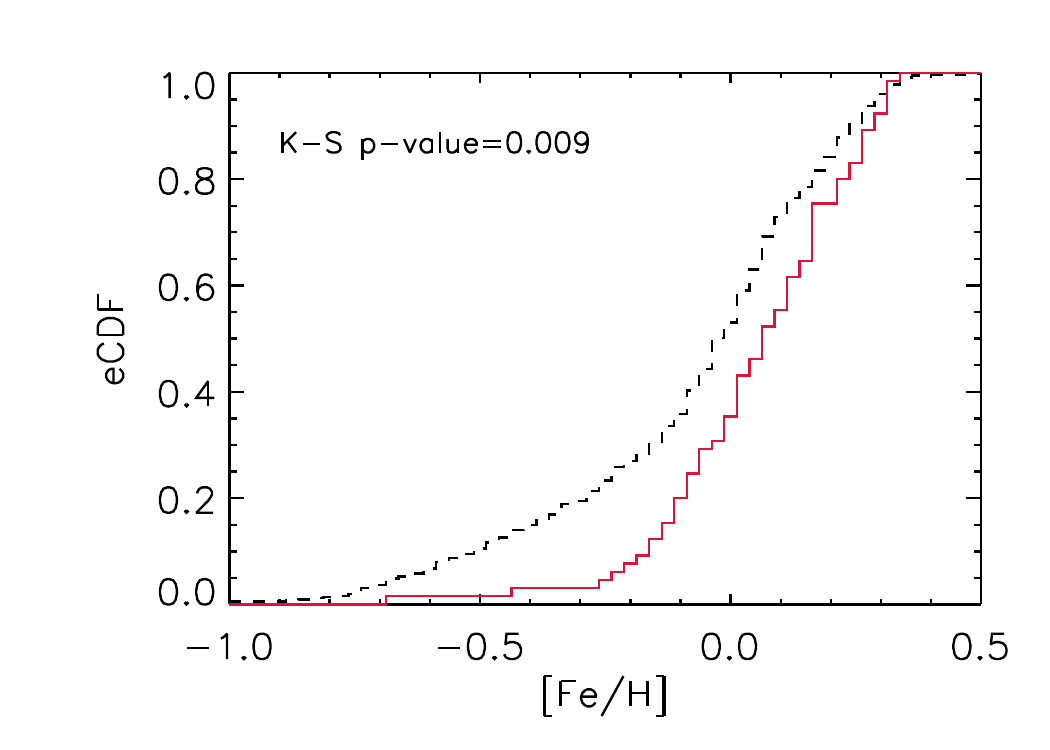} \\
\caption{\emph{Upper panels}: same as left panel of Fig.~\ref{figure_metmstardistr_SP_MegaRV} for the three different bins of stellar mass $0.6-0.8\,\Msun$ (left), $0.8- 1.0\,\Msun$ (center), and $1.0-1.2\,\Msun$ (right). Note that the relative stellar counts of the SP and RV merged samples cannot be directly compared as they were normalized from different distributions. 
\emph{Lower panels:} empirical cumulative distribution functions for the same bins of stellar mass as above. The p-values of the Kolmogorov-Smirnov test are also reported.}
\label{figure_metdistr_massbins_SP_MegaRV}
\end{figure}

\onecolumn

\section{Completenesses of the considered samples}

\begin{figure*}[ht]
    \centering
    \subfigure[SP RV sample]{
        \includegraphics[width=0.325\textwidth]{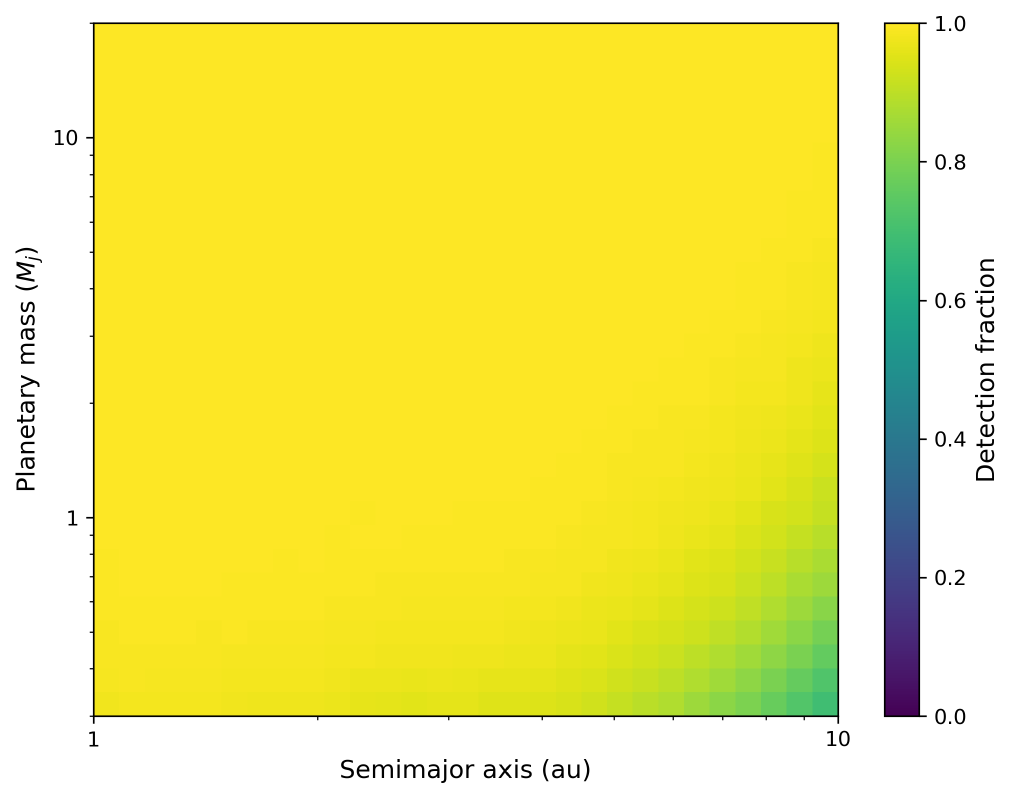}}
    \subfigure[SP transit sample]{
        \includegraphics[width=0.325\textwidth]{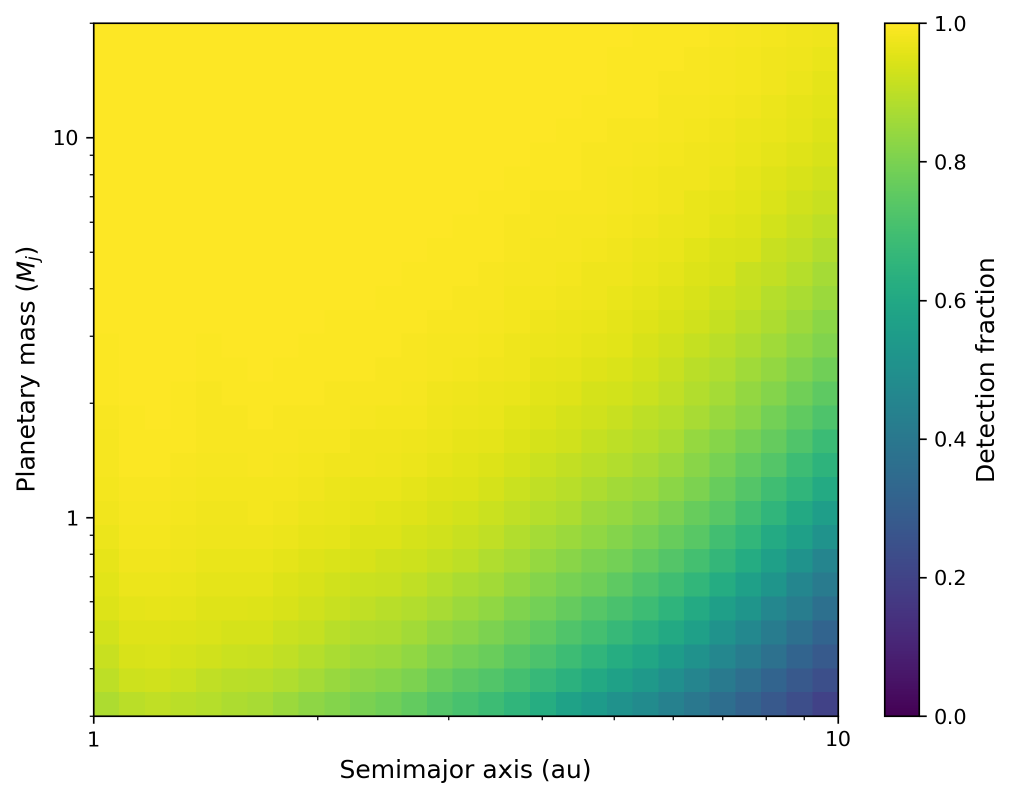}}
    \subfigure[SP (transit+RV) sample]{
        \includegraphics[width=0.325\textwidth]{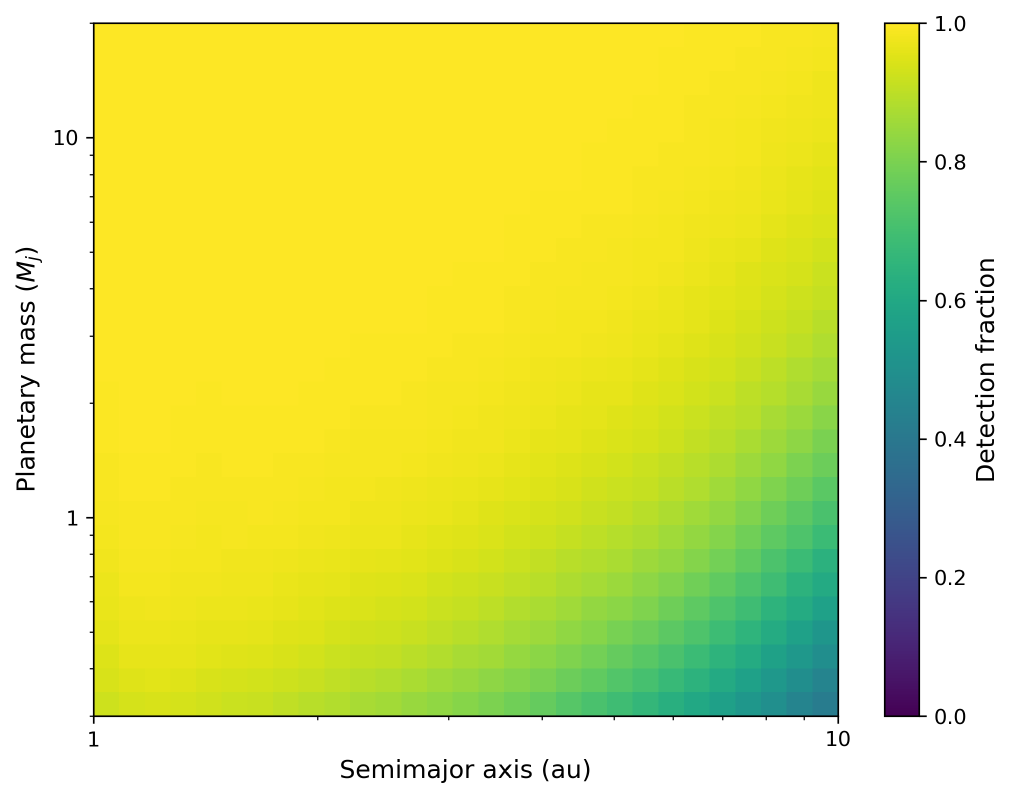}}
    \subfigure[SP and mixed RV sample]{
        \includegraphics[width=0.325\textwidth]{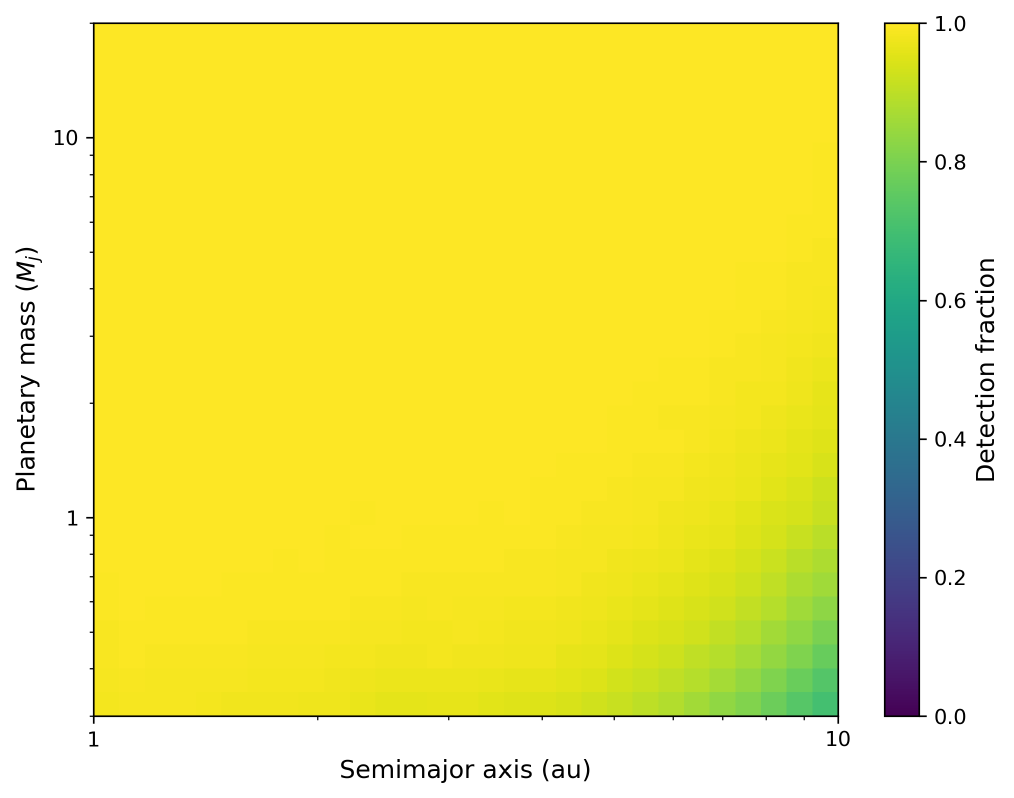}}
    \subfigure[SP and mixed transit sample]{
        \includegraphics[width=0.325\textwidth]{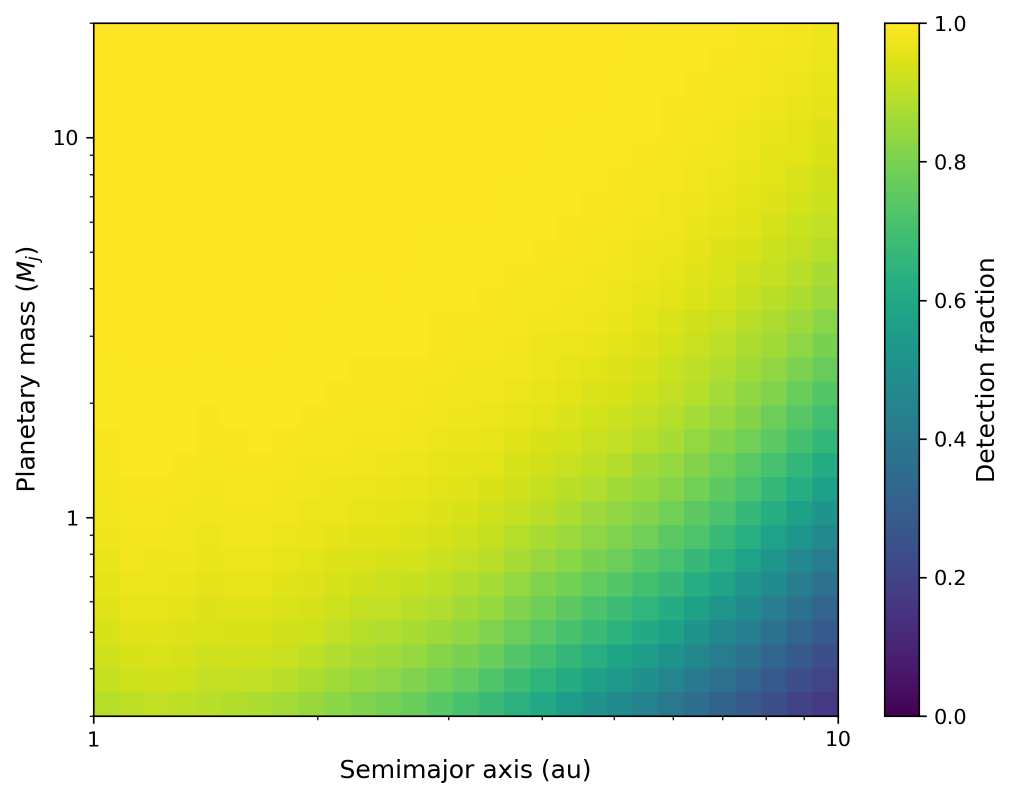}}
    \subfigure[SP and mixed (transit+RV) sample]{
        \includegraphics[width=0.325\textwidth]{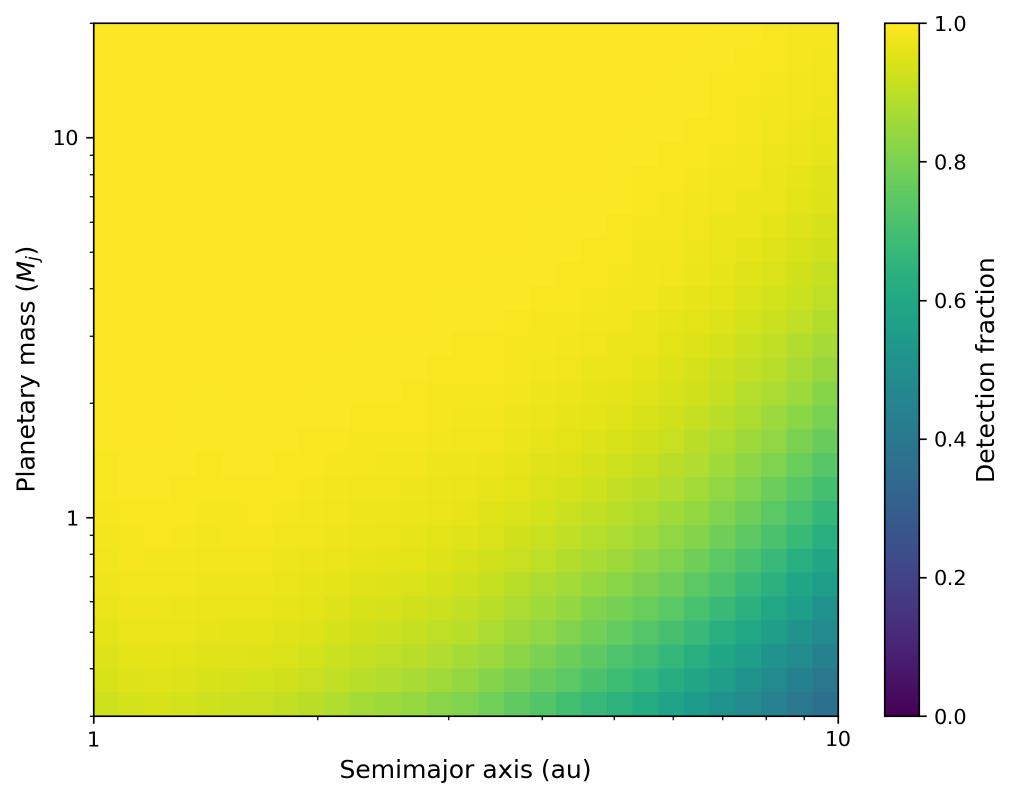}}
    \subfigure[AAT RV survey]{
        \includegraphics[width=0.325\textwidth]{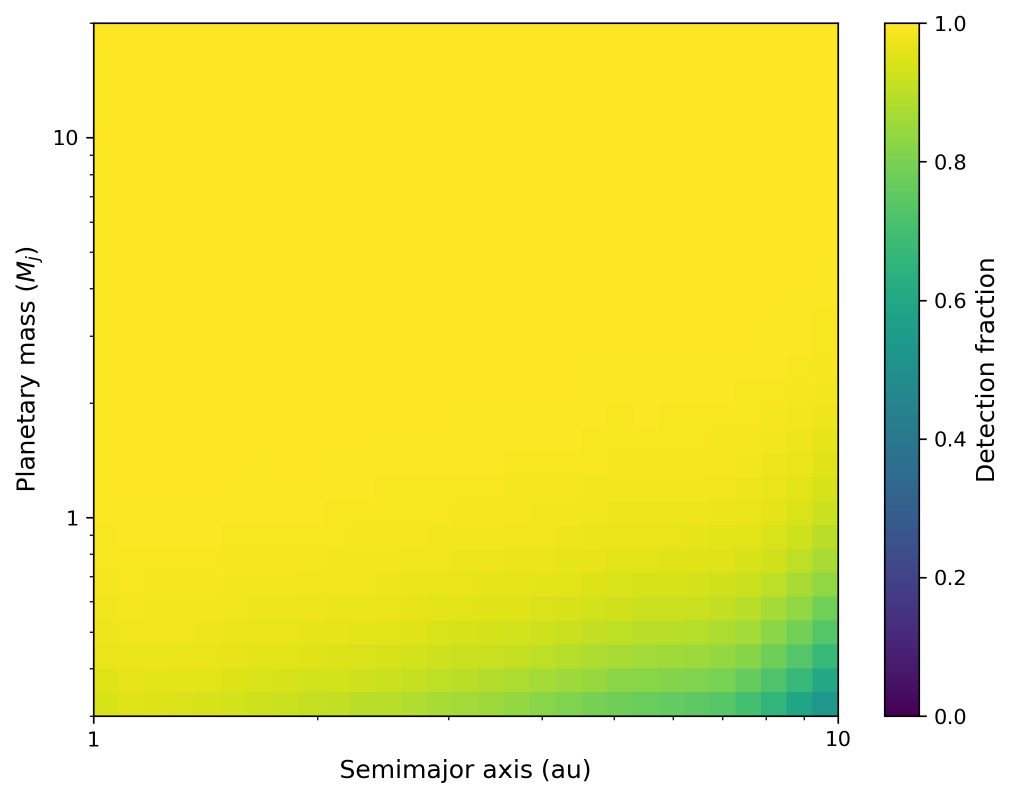}}
    \subfigure[CLS RV survey]{
        \includegraphics[width=0.325\textwidth]{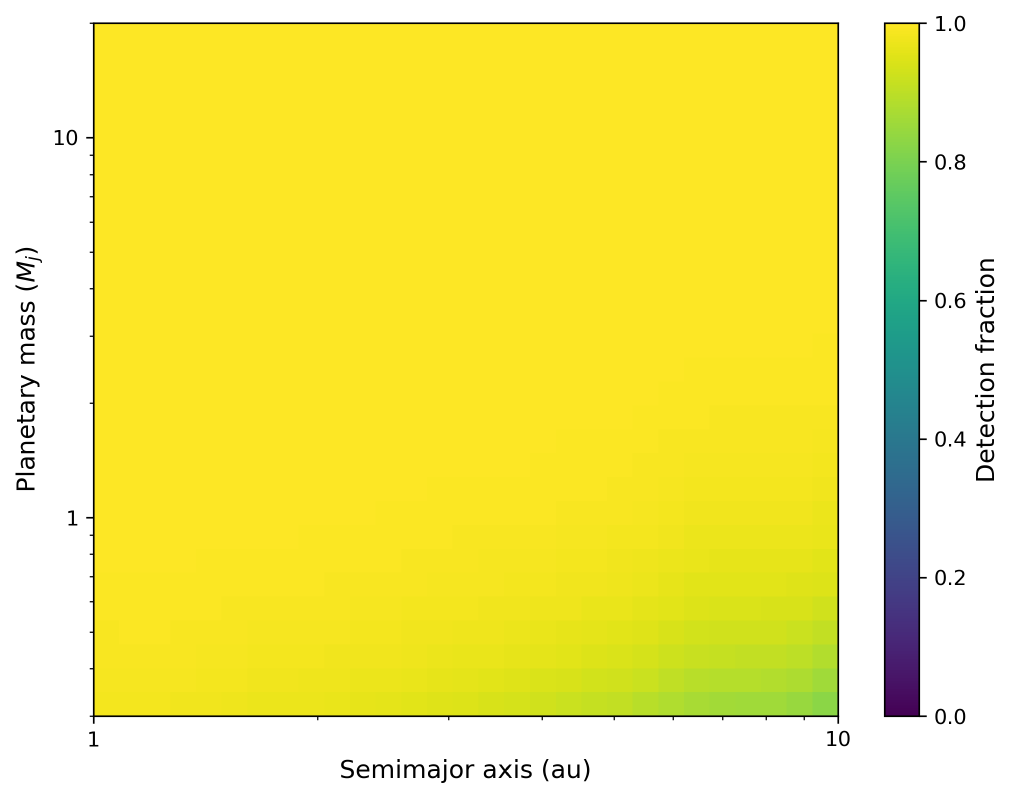}}
    \subfigure[HARPS RV survey]{
        \includegraphics[width=0.325\textwidth]{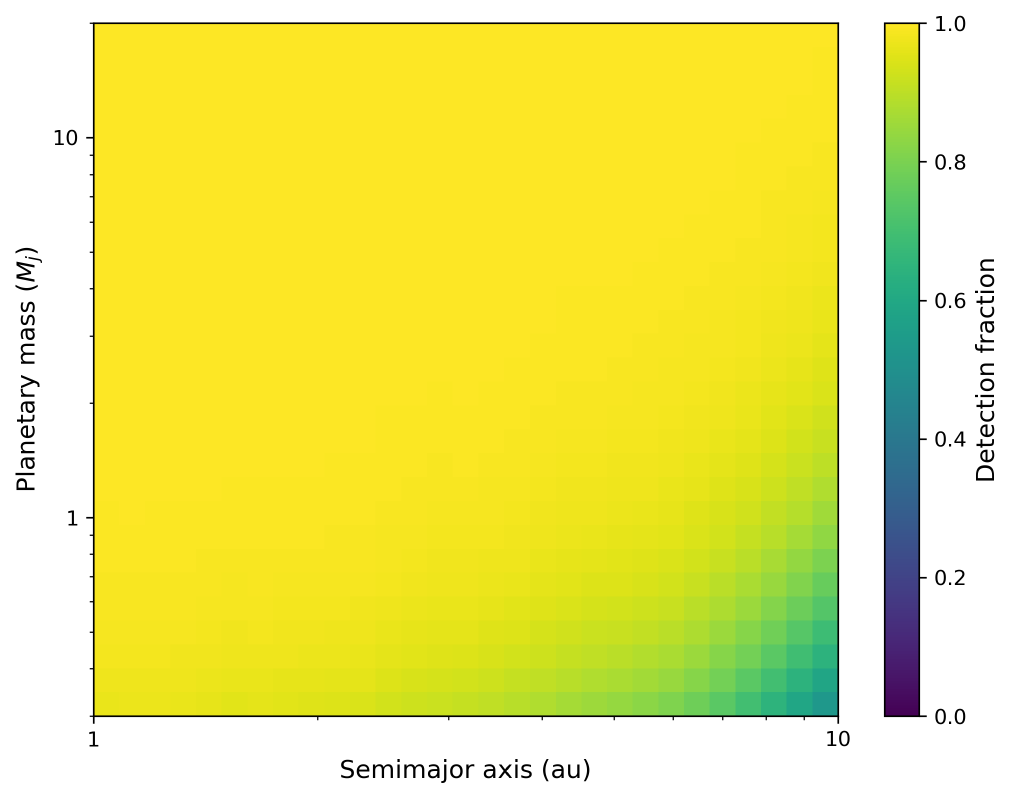}}
    \subfigure[Merged (AAT+CLS+HARPS) RV sample]{
        \includegraphics[width=0.325\textwidth]{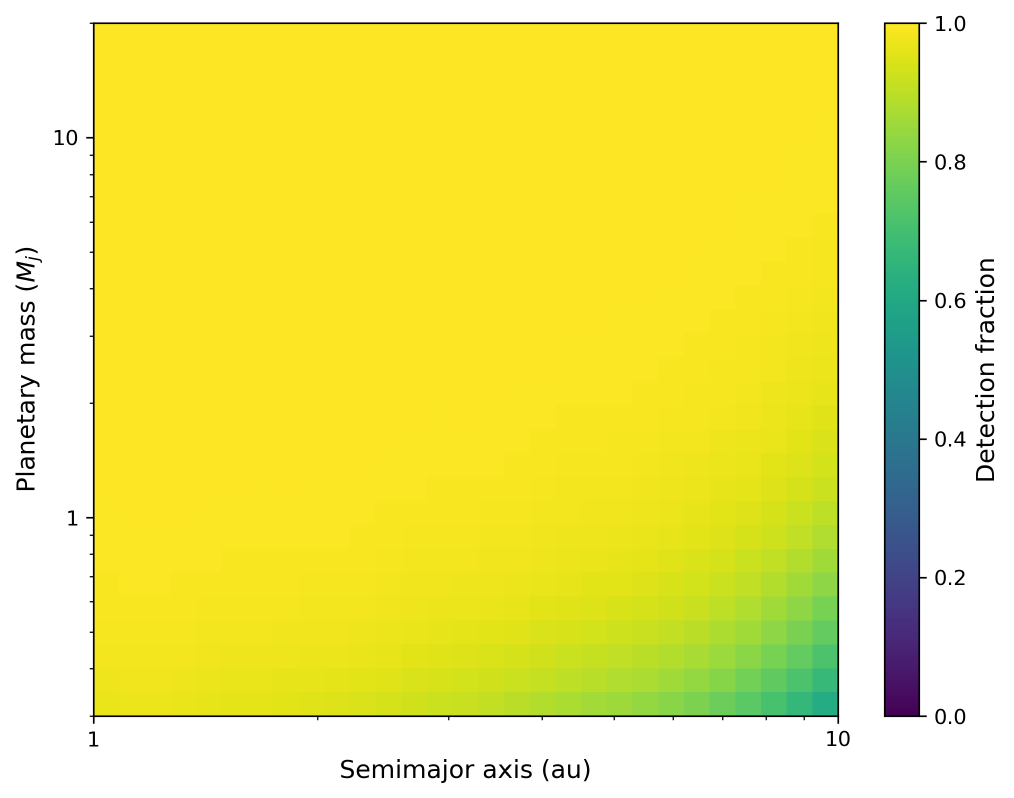}}
    \caption{Average completenesses of all the samples considered in this work. As indicated in the color bars, the detection rate/sensitivity increases from dark blue ($0\%$) to green ($\sim50-80\%$) and yellow ($100\%$).}
    \label{fig:all_completenesses}
\end{figure*}

\clearpage
\newpage

\onecolumn

~\\

\section{Cold Jupiters and their occurrence rates in the small planet and radial velocity comparison samples}

\begin{figure}[h!]
\centering
\vspace{3.0 cm}
\includegraphics[width=1.0\textwidth]{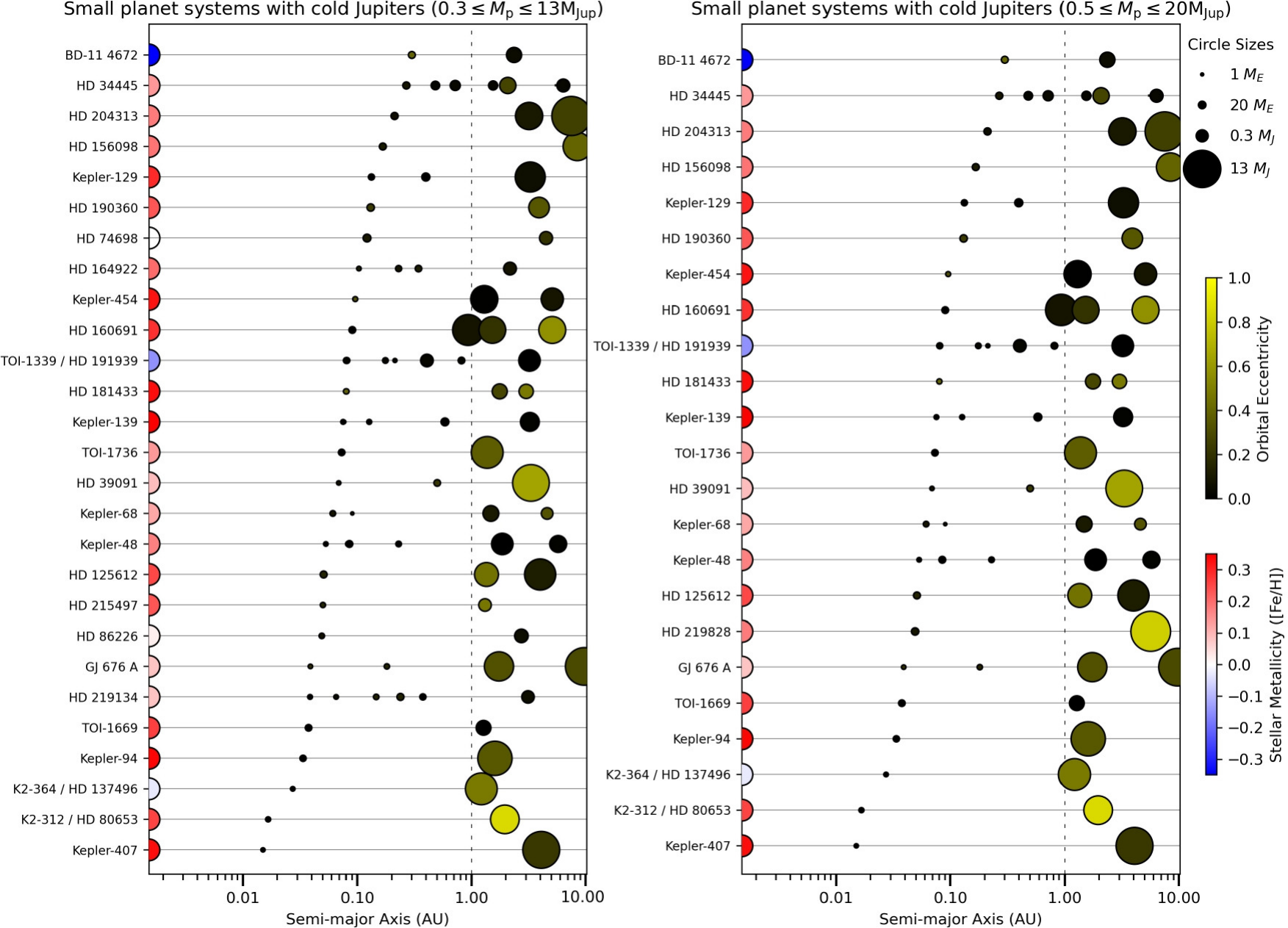}
\caption{The architectures of the systems with both SPs and CJs considered in this work and ordered according to increasing semimajor axis of the innermost planet (from bottom to top). The CJs in the left (27 systems) and right (23 systems) panels have masses $0.3\leq{\rm M_p}\leq13\,\Mjup$ and $0.5\leq{\rm M_p}\leq20\,\Mjup$, respectively, with a large overlap for CJs with $0.5\leq{\rm M_p}\leq13\,\Mjup$. The planet circles have sizes proportional to their mass (see legend in the top right corner), and colors indicating orbital eccentricity (more eccentric planets are shown in yellow). Host stars are color coded as a function of stellar metallicity (metallicity increases from blue to red). }
\label{fig:SP_CJ}
\end{figure}

\clearpage
\newpage


\begin{figure}[h!]
\centering
\includegraphics[width=0.5\textwidth]{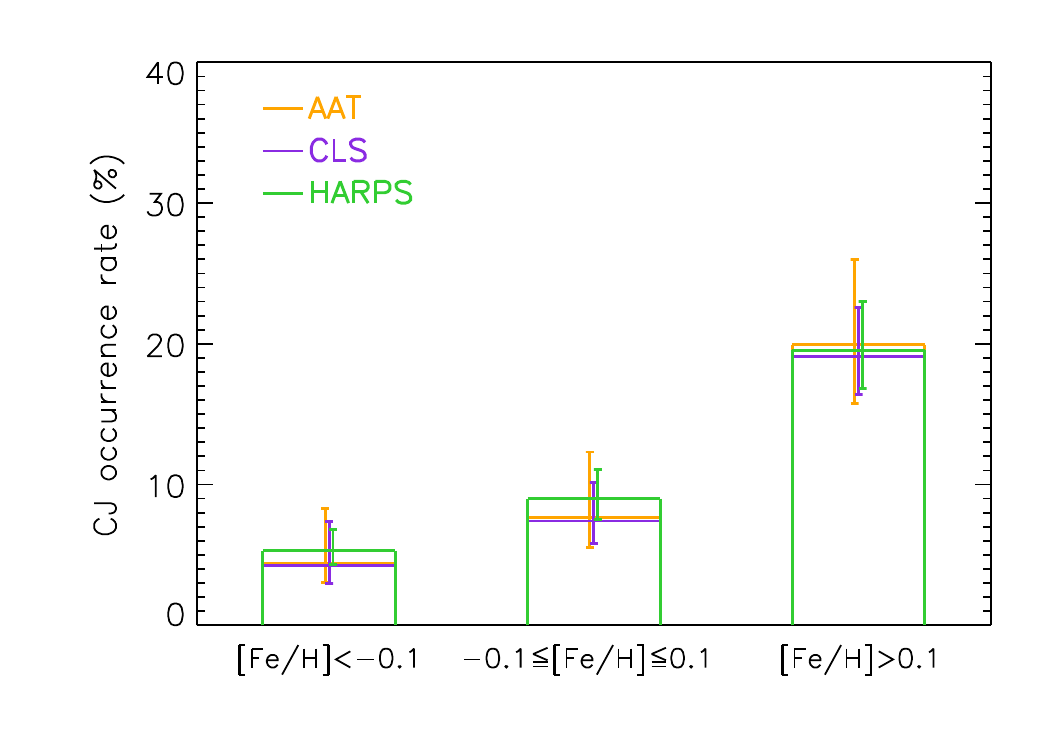}
\caption{Similarly to Fig.~\ref{figure_occurrence_rates_metstar}, occurrence rates of CJs with $\mp=0.5-20\,\Mjup$ in the AAT (orange), CLS (violet), and HARPS (green) RV surveys, at sub-solar ($\rm[Fe/H]<-0.1$), solar ($\rm-0.1\le[Fe/H]\le0.1$), and super-solar ($\rm[Fe/H]>0.1$) metallicity. Almost identical occurrence rates are obtained for $\mp=0.3-13\,\Mjup$. See Table~\ref{table_RVsurveys_occurrence_rates}.}
\label{figure_occurrence_rates_RVsurveys}
\end{figure}

\begin{figure}[h!]
\centering
\includegraphics[width=0.5\textwidth]{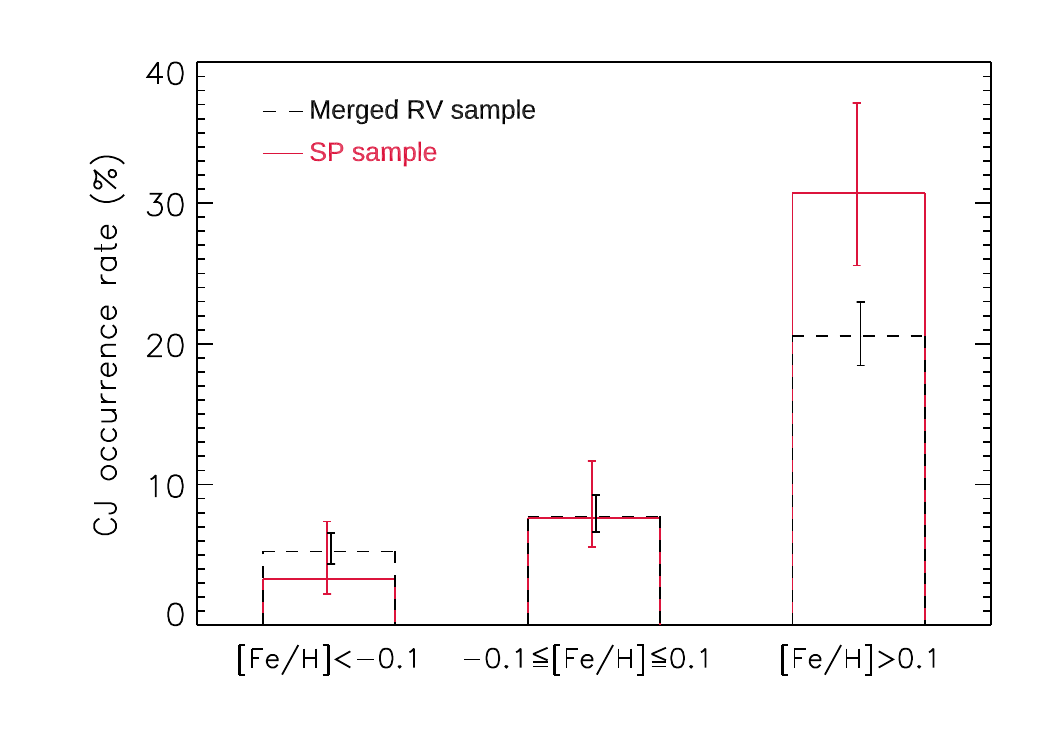}
\caption{Same as Fig.~\ref{figure_occurrence_rates_metstar} for $\mp=0.3-13\,\Mjup$. See also Table~\ref{table_occurrence_rates_metstar}. }
\label{figure_occurrence_rates_metstar_03_13_Mjup}
\end{figure}

\begin{figure}[h!]
\centering
\includegraphics[width=0.350\columnwidth]{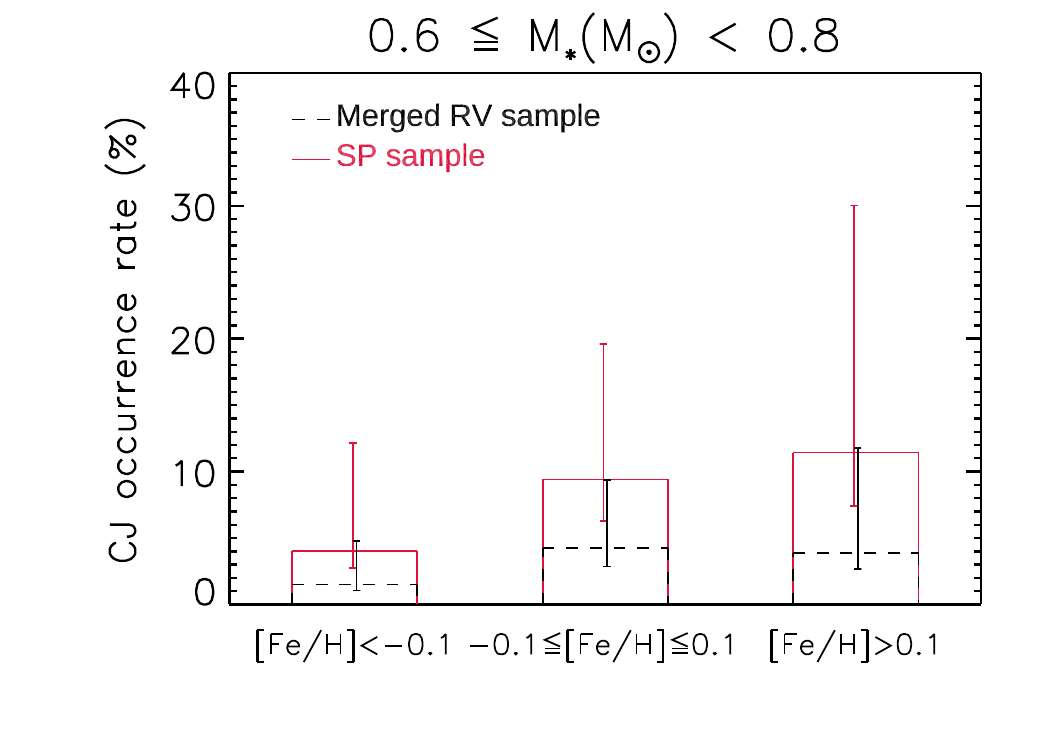}
\hspace{-0.75cm}
\includegraphics[width=0.350\columnwidth]{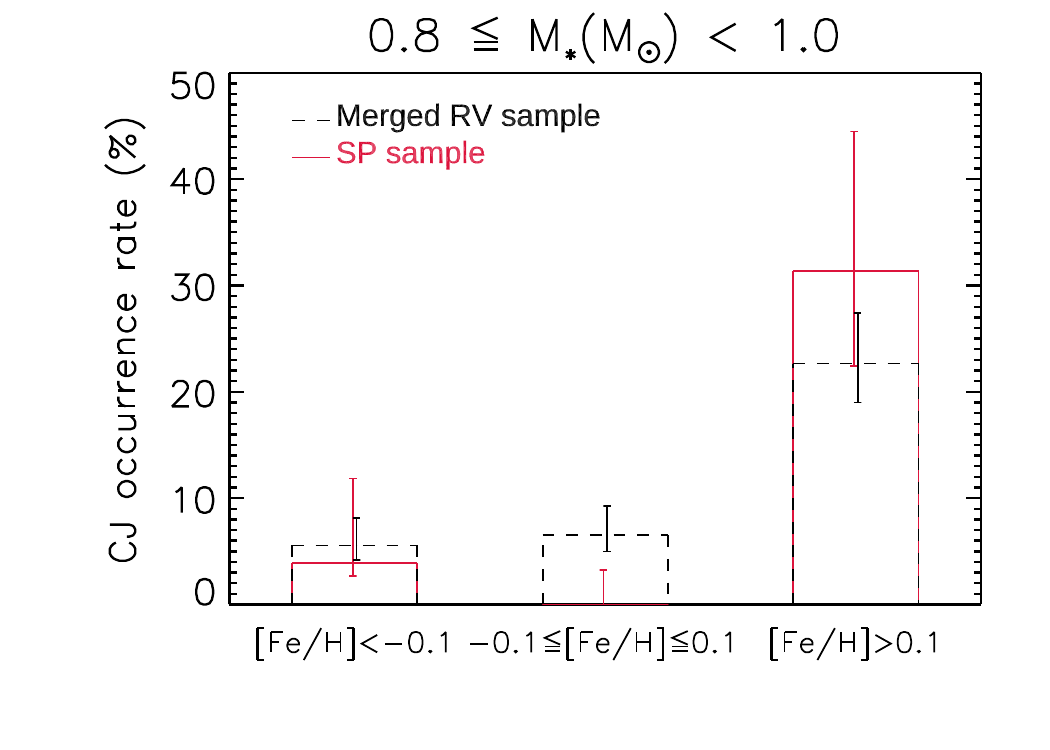}
\hspace{-0.75cm}
\includegraphics[width=0.350\columnwidth]{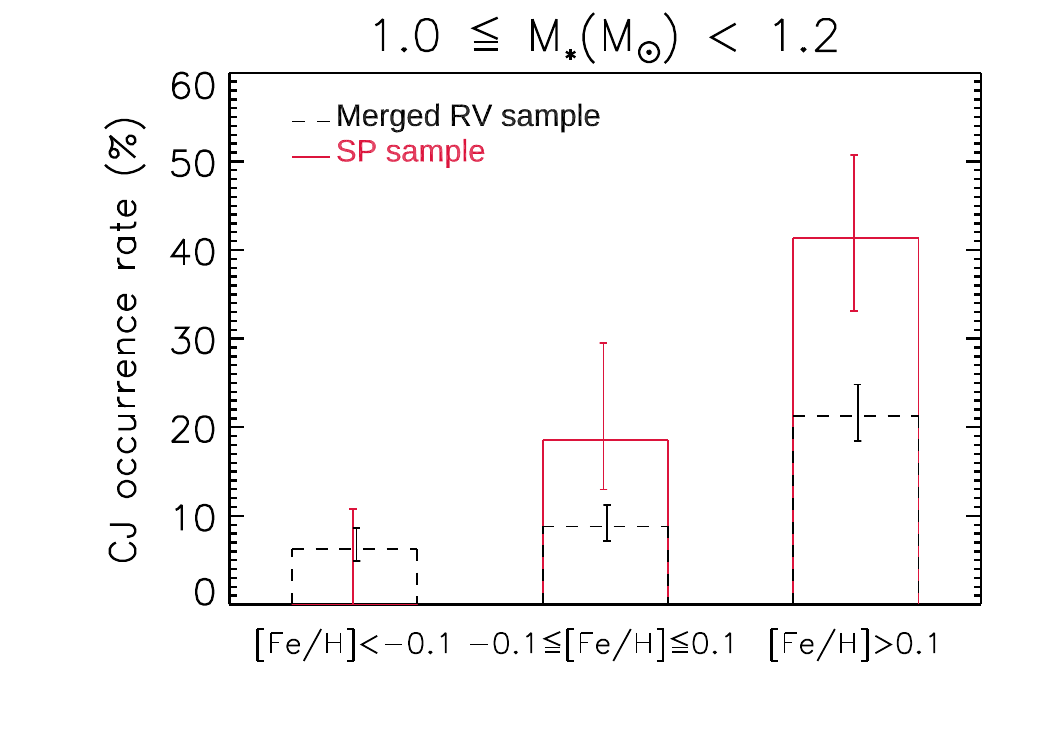} \\
\caption{Same as Fig.~\ref{figure_occurrence_rates_mstar_metstar} for $\mp = 0.3-13\,\Mjup$. See also Table~\ref{table_occurrence_rates_mstar_metstar}. }
\label{figure_occurrence_rates_mstar_metstar_03_13_Mjup}
\end{figure}

\clearpage
\newpage

~\\ 

\section{Occurrence rates of cold Jupiters in the small planet and in the large radial velocity comparison samples as a function of stellar metallicity and/or mass.}

~\\
~\\
~\\
~\\

\renewcommand{\arraystretch}{1.5}

\begin{table*}[h!]
\small
\centering
\caption{Occurrence rates of cold Jupiters in the small planet sample and in the large radial velocity comparison one as a function of stellar metallicity.}   
\begin{tabular}{|c|c|c|c|c|c|c|}
\cline{2-7}
\multicolumn{1}{c|}{} & \multicolumn{3}{c|}{$\mp=0.3-13\,\Mjup$} & \multicolumn{3}{c|}{$\mp=0.5-20\,\Mjup$} \\ 
\hline
\multicolumn{1}{|c|}{[Fe/H]} & \multicolumn{1}{c|}{$N_\star\,|\,d$} & \multicolumn{1}{c|}{$C$} & \multicolumn{1}{c|}{$f_{\rm CJ}$ [\%]} &
\multicolumn{1}{c|}{$N_\star\,|\,d$} & \multicolumn{1}{c|}{$C$} & \multicolumn{1}{c|}{$f_{\rm CJ}$ [\%]} \\  
\hline
\hline
\multicolumn{7}{|c|}{\textbf{\normalsize Small planet (transit+radial velocity) sample}} \\
\hline
\hline
$\overline{\rm [Fe/H]}=-0.011\pm0.005$   & $217\,|\,27$     & 0.928     & $13.4^{+2.8}_{-2.0}$     & $217\,|\,23$     & 0.958     & $11.1^{+2.5}_{-1.8}$ \\
\hline
\cline{1-7}
$\rm [Fe/H]<-0.1$                                       & $64\,|\,2$        & 0.943      & $3.3^{+4.1}_{-1.1}$       & $64\,|\,2$        & 0.969      & $3.2^{+4.0}_{-1.0}$ \\
\hline
$-0.1 \le \rm [Fe/H] \le +0.1$                      & $86\,|\,6$         & 0.918     & $7.6^{+4.1}_{-2.0}$       & $86\,|\,3$         & 0.952      & $3.7^{+3.3}_{-1.1}$ \\
\hline
$\rm [Fe/H]>+0.1$                                      & $67\,|\,19$       & 0.923     & $30.7^{+6.4}_{-5.2}$     & $67\,|\,18$       & 0.954      & $28.2^{+6.2}_{-4.9}$ \\
\hline
\cline{1-7}
$\rm [Fe/H] \leq 0$                                       & $102\,|\,3$         & 0.932      & $3.2^{+2.9}_{-1.0}$       & $102\,|\,3$         & 0.961      & $3.1^{+2.8}_{-0.9}$ \\
\hline
$\rm [Fe/H] > 0$                                       & $115\,|\,24$        & 0.923      & $22.6^{+4.5}_{-3.5}$   &    $115\,|\,20$        & 0.954      & $18.2^{+4.2}_{-3.1}$ \\
\hline
\hline
\multicolumn{7}{|c|}{\textbf{\normalsize Merged (AAT+CLS+HARPS) radial velocity comparison sample}} \\
\hline
\hline
$\overline{\rm [Fe/H]}=-0.072\pm0.009$   & $1167\,|\,118$     & 0.974     & $10.4^{+1.0}_{-0.8}$     & $1167\,|\,113$     & 0.988     & $9.8^{+0.9}_{-0.8}$ \\
\hline
\cline{1-7}
$\rm [Fe/H]<-0.1$                                       & $430\,|\,22$        & 0.974      & $5.2^{+1.3}_{-0.9}$       & $430\,|\,21$        & 0.988      & $4.9^{+1.3}_{-0.8}$ \\
\hline
$-0.1 \le \rm [Fe/H] \le +0.1$                      & $413\,|\,31$         & 0.971     & $7.7^{+1.5}_{-1.1}$       & $413\,|\,31$         & 0.986      & $7.6^{+1.5}_{-1.1}$ \\
\hline
$\rm [Fe/H]>+0.1$                                      & $324\,|\,65$        & 0.977     & $20.5^{+2.4}_{-2.1}$      & $324\,|\,61$          & 0.990      & $19.0^{+2.4}_{-2.0}$ \\
\hline
\cline{1-7}
$\rm [Fe/H] \leq 0$                                       & $624\,|\,33$         & 0.973      & $5.4^{+1.1}_{-0.8}$       & $624\,|\,32$         & 0.988      & $5.2^{+1.0}_{-0.7}$ \\
\hline
$\rm [Fe/H] > 0$                                       & $543\,|\,85$        & 0.974      & $16.1^{+1.7}_{-1.5}$   &    $543\,|\,81$        & 0.988      & $15.1^{+1.7}_{-1.4}$ \\
\hline
\end{tabular}
\tablefoot{From left to right, the columns report the stellar metallicity ([Fe/H]), the total number of stars ($N_\star$) and the number of stars with confirmed CJs ($d$), 
the average completeness ($C$), and the occurrence rate of CJs ($f_{\rm CJ}$), for the SP (top) and RV comparison sample (bottom) and their subsamples at 
sub-solar, solar, and super-solar metallicity, by considering two ranges for the CJ mass: $0.3-13\,\Mjup$ (columns 2-4) and $0.5-20\,\Mjup$ (columns 5-7). We note that $f_{\rm CJ}$ for the SP sample is equivalent to $f_{\rm CJ|SP}$.}
\label{table_occurrence_rates_metstar}
\end{table*}

\clearpage
\newpage

~\\
~\\

\renewcommand{\arraystretch}{1.5}

\begin{table*}[h]
\tiny
\centering
\caption{Occurrence rates of cold Jupiters in the small planet sample and in the large radial velocity comparison one as a function of stellar metallicity in three intervals of stellar mass.}   
\begin{tabular}{|c|c|c|c|c|c|c|}
\cline{2-7}
\multicolumn{1}{c|}{} & \multicolumn{3}{c|}{$\mp=0.3-13\,\Mjup$} & \multicolumn{3}{c|}{$\mp=0.5-20\,\Mjup$} \\ 
\hline
\multicolumn{1}{|c|}{[Fe/H]} & \multicolumn{1}{c|}{$N_\star\,|\,d$} & \multicolumn{1}{c|}{$C$} & \multicolumn{1}{c|}{$f_{\rm CJ}$ [\%]} &
\multicolumn{1}{c|}{$N_\star\,|\,d$} & \multicolumn{1}{c|}{$C$} & \multicolumn{1}{c|}{$f_{\rm CJ}$ [\%]} \\  
\hline
\hline
\hline
\multicolumn{7}{|c|}{\textbf{\large $0.6 \leq M_{\star} < 0.8\, \rm M_\odot$}} \\
\hline
\cline{1-7}
\multicolumn{7}{|c|}{\textbf{\normalsize Small planet (transit+radial velocity) sample}} \\
\hline 
\cline{1-7}
$\overline{\rm [Fe/H]}=-0.114\pm0.011$   & $58\,|\,4$     & 0.947     & $7.3^{+5.1}_{-2.1}$     & $58\,|\,3$     & 0.971     & $5.3^{+4.7}_{-1.6}$ \\
\hline
\cline{1-7}
$\rm [Fe/H]<-0.1$                                       & $26\,|\,1$         & 0.959      & $4.0^{+8.1}_{-1.2}$       & $26\,|\,1$        & 0.979      & $3.9^{+8.0}_{-1.2}$ \\
\hline
$-0.1 \le \rm [Fe/H] \le +0.1$                      & $23\,|\,2$         & 0.924     & $9.4^{+10.2}_{-3.1}$       & $23\,|\,1$         & 0.955     & $4.5^{+9.1}_{-1.4}$ \\
\hline
$\rm [Fe/H]>+0.1$                                      & $9\,|\,1$         & 0.972     & $11.4^{+18.6}_{-4.0}$     & $9\,|\,1$          & 0.986      & $11.3^{+18.4}_{-3.9}$ \\
\hline
\cline{1-7}
\multicolumn{7}{|c|}{\textbf{\normalsize Merged (AAT+CLS+HARPS) radial velocity comparison sample}} \\
\hline
\cline{1-7}
$\overline{\rm [Fe/H]}=-0.121\pm0.014$   & $142\,|\,4$     & 0.983     & $2.9^{+2.2}_{-0.8}$     & $142\,|\,4$     & 0.993     & $2.8^{+2.1}_{-0.9}$ \\
\hline
\cline{1-7}
$\rm [Fe/H]<-0.1$                                       & $68\,|\,1$         & 0.980      & $1.5^{+3.3}_{-0.4}$       & $68\,|\,1$        & 0.991      & $1.5^{+3.2}_{-0.4}$ \\
\hline
$-0.1 \le \rm [Fe/H] \le +0.1$                      & $48\,|\,2$         & 0.984     & $4.2^{+5.1}_{-1.4}$       & $48\,|\,2$         & 0.993     & $4.2^{+5.1}_{-1.4}$ \\
\hline
$\rm [Fe/H]>+0.1$   &          $26\,|\,1$         & 0.991      & $3.9^{+7.9}_{-1.2}$       & $26\,|\,1$        & 0.997      & $3.9^{+7.8}_{-1.2}$ \\
\hline
\hline
\hline
\multicolumn{7}{|c|}{\textbf{\large $0.8 \leq M_{\star} < 1.0\, \rm M_\odot$}} \\
\hline
\cline{1-7}
\multicolumn{7}{|c|}{\textbf{\normalsize Small planet (transit+radial velocity) sample}} \\
\hline 
\cline{1-7}
$\overline{\rm [Fe/H]}=-0.030\pm0.007$   & $82\,|\,6$     & 0.923     & $7.9^{+4.3}_{-2.1}$     & $82\,|\,4$     & 0.954     & $5.1^{+3.7}_{-1.5}$ \\
\hline
\cline{1-7}
$\rm [Fe/H]<-0.1$                                       & $27\,|\,1$         & 0.948      & $3.9^{+7.9}_{-1.2}$       & $27\,|\,1$        & 0.971      & $3.8^{+7.8}_{-1.2}$ \\
\hline
$-0.1 \le \rm [Fe/H] \le +0.1$                      & $37\,|\,0$         & 0.922     & $< 3.2$       & $37\,|\,0$         & 0.955     & $< 3.1$ \\
\hline
$\rm [Fe/H]>+0.1$                                      & $18\,|\,5$         & 0.886     & $31.3^{+13.1}_{-8.9}$     & $18\,|\,3$          & 0.926      & $18.0^{+12.7}_{-5.8}$ \\
\hline
\cline{1-7}
\multicolumn{7}{|c|}{\textbf{\normalsize Merged (AAT+CLS+HARPS) radial velocity comparison sample}} \\
\hline
\cline{1-7}
$\overline{\rm [Fe/H]}=-0.073\pm0.016$   & $387\,|\,39$     & 0.981     & $10.3^{+1.8}_{-1.3}$     & $387\,|\,36$     & 0.992     & $9.4^{+1.7}_{-1.3}$ \\
\hline
\cline{1-7}
$\rm [Fe/H]<-0.1$                                       & $147\,|\,8$         & 0.981      & $5.5^{+2.6}_{-1.3}$       & $147\,|\,8$        & 0.992      & $5.5^{+2.5}_{-1.3}$ \\
\hline
$-0.1 \le \rm [Fe/H] \le +0.1$                      & $141\,|\,9$         & 0.981     & $6.5^{+2.7}_{-1.5}$       & $141\,|\,9$         & 0.991     & $6.4^{+2.7}_{-1.5}$ \\
\hline
$\rm [Fe/H]>+0.1$   &          $99\,|\,22$         & 0.981      & $22.6^{+4.8}_{-3.7}$       & $99\,|\,19$        & 0.992      & $19.3^{+4.6}_{-1.3}$ \\
\hline
\hline
\hline
\multicolumn{7}{|c|}{\textbf{\large $1.0 \leq M_{\star} < 1.2\, \rm M_\odot$}} \\
\hline
\cline{1-7}
\multicolumn{7}{|c|}{\textbf{\normalsize Small planet (transit+radial velocity) sample}} \\
\hline 
\cline{1-7}
$\overline{\rm [Fe/H]}=0.077\pm0.008$   & $65\,|\,16$     & 0.915     & $26.9^{+6.4}_{-4.9}$     & $65\,|\,15$     & 0.951     & $24.3^{+6.2}_{-4.6}$ \\
\hline
\cline{1-7}
$\rm [Fe/H]<-0.1$                                       & $10\,|\,0$         & 0.903      & $<10.8$       & $10\,|\,0$        & 0.950      & $< 10.3$ \\
\hline
$-0.1 \le \rm [Fe/H] \le +0.1$                      & $24\,|\,4$         & 0.899     & $18.5^{+10.9}_{-5.6}$       & $24\,|\,2$         & 0.941     & $8.9^{+9.7}_{-3.0}$ \\
\hline
$\rm [Fe/H]>+0.1$                                      & $31\,|\,12$         & 0.936     & $41.4^{+9.4}_{-8.2}$     & $31\,|\,13$          & 0.959      & $43.7^{+9.2}_{-8.4}$ \\
\hline
\cline{1-7}
\multicolumn{7}{|c|}{\textbf{\normalsize Merged (AAT+CLS+HARPS) radial velocity comparison sample}} \\
\hline
\cline{1-7}
$\overline{\rm [Fe/H]}=-0.056\pm0.014$   & $549\,|\,63$     & 0.972     & $11.8^{+1.5}_{-1.3}$     & $549\,|\,62$     & 0.988     & $11.4^{+1.5}_{-1.2}$ \\
\hline
\cline{1-7}
$\rm [Fe/H]<-0.1$                                       & $181\,|\,11$         & 0.973      & $6.2^{+2.3}_{-1.4}$       & $181\,|\,10$        & 0.988      & $5.6^{+2.2}_{-1.3}$ \\
\hline
$-0.1 \le \rm [Fe/H] \le +0.1$                      & $200\,|\,17$         & 0.966     & $8.8^{+2.5}_{-1.6}$       & $200\,|\,17$         & 0.984     & $8.6^{+2.4}_{-1.6}$ \\
\hline
$\rm [Fe/H]>+0.1$   &          $168\,|\,35$         & 0.978      & $21.3^{+3.5}_{-2.8}$       & $168\,|\,35$        & 0.991      & $21.0^{+3.5}_{-2.8}$ \\
\hline
\end{tabular}
\tablefoot{Same columns as Table~\ref{table_occurrence_rates_metstar}.}
\label{table_occurrence_rates_mstar_metstar}
\end{table*}

\clearpage
\newpage

\section{Occurrence rates of cold Jupiters in the small planet transit, radial velocity, and transit+radial velocity samples, without and with mixed systems, as a function of stellar metallicity.}

\renewcommand{\arraystretch}{1.3}

\begin{table*}[h]
\tiny
\centering
\caption{Occurrence rates of cold Jupiters in the small planet transit, radial velocity, and transit+radial velocity samples, without and with the inclusion of mixed systems, as a function of stellar metallicity.}   
\begin{tabular}{|c|c|c|c|c|c|c|}
\cline{2-7}
\multicolumn{1}{c|}{} & \multicolumn{3}{c|}{$\mp=0.3-13\,\Mjup$} & \multicolumn{3}{c|}{$\mp=0.5-20\,\Mjup$} \\ 
\hline
\multicolumn{1}{|c|}{[Fe/H]} & \multicolumn{1}{c|}{$N_\star\,|\,d$} & \multicolumn{1}{c|}{$C$} & \multicolumn{1}{c|}{$f_{\rm CJ}$ [\%]} &
\multicolumn{1}{c|}{$N_\star\,|\,d$} & \multicolumn{1}{c|}{$C$} & \multicolumn{1}{c|}{$f_{\rm CJ}$ [\%]} \\  
\hline
\hline
\multicolumn{7}{|c|}{\textbf{\normalsize Small planet transit sample}} \\
\hline
\hline
$\overline{\rm [Fe/H]}=+0.002\pm0.006$   & $134\,|\,12$     & 0.894     & $10.0^{+3.4}_{-2.1}$     & $134\,|\,12$     & 0.938     & $9.5^{+3.3}_{-2.0}$ \\
\hline 
\cline{1-7} 
$\rm [Fe/H]<-0.1$                                       & $36\,|\,1$         & 0.910      & $3.0^{+6.4}_{-0.9}$       & $36\,|\,1$        & 0.949      & $2.9^{+6.1}_{-0.9}$ \\
\hline
$-0.1 \le \rm [Fe/H] \le +0.1$                      & $56\,|\,1$         & 0.884     & $2.0^{+4.3}_{-0.6}$       & $56\,|\,1$         & 0.931     & $1.9^{+4.1}_{-0.6}$ \\
\hline
$\rm [Fe/H]>+0.1$                                      & $42\,|\,10$         & 0.889     & $26.8^{+8.3}_{-5.9}$     & $42\,|\,10$          & 0.932      & $25.5^{+8.0}_{-5.6}$ \\
\hline
\cline{1-7}
$\rm [Fe/H] \leq 0$                                       & $62\,|\,2$         & 0.900      & $3.6^{+4.4}_{-1.1}$       & $62\,|\,2$         & 0.942      & $3.4^{+4.2}_{-1.1}$ \\
\hline
$\rm [Fe/H] > 0$                                       & $72\,|\,10$        & 0.886      & $15.7^{+5.6}_{-3.5}$   &    $72\,|\,10$        & 0.931      & $14.9^{+5.4}_{-3.3}$ \\
\hline
\hline
\multicolumn{7}{|c|}{\textbf{\normalsize Small planet radial velocity sample}} \\
\hline
\hline
$\overline{\rm [Fe/H]}=-0.033\pm0.008$   & $83\,|\,15$       & 0.982     & $18.4^{+5.0}_{-3.5}$     & $83\,|\,11$     & 0.992     & $13.4^{+4.6}_{-2.9}$ \\
\hline
\cline{1-7}
$\rm [Fe/H]<-0.1$                                       & $28\,|\,1$        & 0.985      & $3.6^{+7.4}_{-1.1}$       & $28\,|\,1$       & 0.994     & $3.6^{+7.4}_{-1.1}$ \\
\hline
$-0.1 \le \rm [Fe/H] \le +0.1$                      & $30\,|\,5$         & 0.979     & $17.0^{+9.0}_{-4.8}$     & $30\,|\,2$       & 0.990     & $6.7^{+7.7}_{-2.2}$ \\
\hline
$\rm [Fe/H]>+0.1$                                      & $25\,|\,9$        & 0.982     & $36.7^{+10.3}_{-8.3}$     & $25\,|\,8$       & 0.992    & $32.3^{+10.4}_{-7.7}$ \\
\hline
\cline{1-7}
$\rm [Fe/H] \leq 0$                                       & $40\,|\,1$         & 0.981      & $2.5^{+5.4}_{-0.8}$       & $40\,|\,1$         & 0.991      & $2.5^{+5.3}_{-0.8}$ \\
\hline
$\rm [Fe/H] > 0$                                       & $43\,|\,14$        & 0.983      & $33.1^{+7.8}_{-6.3}$   &    $43\,|\,10$        & 0.992      & $23.4^{+7.6}_{-5.2}$ \\
\hline
\hline
\multicolumn{7}{|c|}{\textbf{\normalsize Small planet (transit+radial velocity) sample}} \\
\hline
\hline
$\overline{\rm [Fe/H]}=-0.011\pm0.005$   & $217\,|\,27$     & 0.928     & $13.4^{+2.8}_{-2.0}$     & $217\,|\,23$     & 0.958     & $11.1^{+2.5}_{-1.8}$ \\
\hline
\cline{1-7}
$\rm [Fe/H]<-0.1$                                       & $64\,|\,2$        & 0.943      & $3.3^{+4.1}_{-1.1}$       & $64\,|\,2$        & 0.969      & $3.2^{+4.0}_{-1.0}$ \\
\hline
$-0.1 \le \rm [Fe/H] \le +0.1$                      & $86\,|\,6$         & 0.918     & $7.6^{+4.1}_{-2.0}$       & $86\,|\,3$         & 0.952      & $3.7^{+3.3}_{-1.1}$ \\
\hline
$\rm [Fe/H]>+0.1$                                      & $67\,|\,19$       & 0.923     & $30.7^{+6.4}_{-5.2}$     & $67\,|\,18$       & 0.954      & $28.2^{+6.2}_{-4.9}$ \\
\hline
\cline{1-7}
$\rm [Fe/H] \leq 0$                                       & $102\,|\,3$         & 0.932      & $3.2^{+2.9}_{-1.0}$       & $102\,|\,3$         & 0.961      & $3.1^{+2.8}_{-0.9}$ \\
\hline
$\rm [Fe/H] > 0$                                       & $115\,|\,24$        & 0.923      & $22.6^{+4.5}_{-3.5}$   &    $115\,|\,20$        & 0.954      & $18.2^{+4.2}_{-3.1}$ \\
\hline
\hline
\multicolumn{7}{|c|}{\textbf{\normalsize Small planet and mixed transit sample}} \\
\hline
\hline
$\overline{\rm [Fe/H]}=+0.022\pm0.006$   & $151\,|\,16$     & 0.891     & $11.9^{+3.4}_{-2.2}$     & $151\,|\,16$     & 0.935     & $11.3^{+3.2}_{-2.1}$ \\
\hline
\cline{1-7}
$\rm [Fe/H]<-0.1$                                       & $36\,|\,1$         & 0.910      & $3.0^{+6.4}_{-0.9}$       & $36\,|\,1$        & 0.949      & $2.9^{+6.1}_{-0.9}$ \\
\hline
$-0.1 \le \rm [Fe/H] \le +0.1$                      & $63\,|\,1$         & 0.881     & $1.8^{+3.9}_{-0.5}$       & $63\,|\,1$         & 0.929     & $1.7^{+3.7}_{-0.5}$ \\
\hline
$\rm [Fe/H]>+0.1$                                      & $52\,|\,14$         & 0.883     & $30.5^{+7.5}_{-5.8}$     & $52\,|\,14$          & 0.928      & $29.0^{+7.2}_{-5.6}$ \\
\hline
\cline{1-7}
$\rm [Fe/H] \leq 0$                                       & $64\,|\,2$         & 0.899      & $3.5^{+4.2}_{-1.1}$       & $64\,|\,2$         & 0.942      & $3.3^{+4.1}_{-1.1}$ \\
\hline
$\rm [Fe/H] > 0$                                       & $87\,|\,14$        & 0.881      & $18.3^{+5.2}_{-3.6}$   &    $87\,|\,14$        & 0.927      & $17.4^{+5.0}_{-3.4}$ \\
\hline
\hline
\multicolumn{7}{|c|}{\textbf{\normalsize Small planet and mixed radial velocity sample}} \\
\hline
\hline
$\overline{\rm [Fe/H]}=-0.018\pm0.007$   & $91\,|\,17$       & 0.983     & $19.0^{+4.8}_{-3.4}$     & $91\,|\,13$     & 0.992     & $14.4^{+4.5}_{-2.9}$ \\
\hline
\cline{1-7}
$\rm [Fe/H]<-0.1$                                       & $29\,|\,1$        & 0.985      & $3.5^{+7.2}_{-1.1}$       & $29\,|\,1$       & 0.993     & $3.5^{+7.2}_{-1.1}$ \\
\hline
$-0.1 \le \rm [Fe/H] \le +0.1$                      & $34\,|\,6$         & 0.979     & $18.0^{+8.4}_{-4.8}$     & $34\,|\,3$       & 0.990     & $8.9^{+7.4}_{-2.8}$ \\
\hline
$\rm [Fe/H]>+0.1$                                      & $28\,|\,10$        & 0.983     & $36.3^{+9.8}_{-7.9}$     & $28\,|\,9$       & 0.993    & $32.4^{+9.8}_{-7.4}$ \\
\hline
\cline{1-7}
$\rm [Fe/H] \leq 0$                                       & $41\,|\,1$         & 0.981      & $2.5^{+5.3}_{-0.8}$       & $41\,|\,1$         & 0.991      & $2.5^{+5.2}_{-0.7}$ \\
\hline
$\rm [Fe/H] > 0$                                       & $50\,|\,16$        & 0.983      & $32.5^{+7.2}_{-5.9}$   &    $50\,|\,12$        & 0.993      & $24.2^{+7.0}_{-5.0}$ \\
\hline
\hline
\multicolumn{7}{|c|}{\textbf{\normalsize Small planet and mixed (transit+radial velocity) sample}} \\
\hline
\hline
$\overline{\rm [Fe/H]}=0.007\pm0.004$   & $242\,|\,33$     & 0.926     & $14.7^{+2.7}_{-2.1}$     & $242\,|\,29$     & 0.957     & $12.5^{+2.5}_{-1.8}$ \\
\hline
\cline{1-7}
$\rm [Fe/H]<-0.1$                                       & $65\,|\,2$        & 0.944      & $3.3^{+4.0}_{-1.0}$       & $65\,|\,2$        & 0.969      & $3.2^{+3.9}_{-1.0}$ \\
\hline
$-0.1 \le \rm [Fe/H] \le +0.1$                      & $97\,|\,7$         & 0.915     & $7.9^{+3.8}_{-2.0}$       & $97\,|\,4$         & 0.950      & $4.3^{+3.2}_{-1.3}$ \\
\hline
$\rm [Fe/H]>+0.1$                                      & $80\,|\,24$       & 0.918     & $32.7^{+5.8}_{-4.9}$     & $80\,|\,23$       & 0.951      & $30.2^{+5.7}_{-4.7}$ \\
\hline
\cline{1-7}
$\rm [Fe/H] \leq 0$                                       & $105\,|\,3$         & 0.931      & $3.1^{+2.8}_{-0.9}$       & $105\,|\,3$         & 0.961      & $3.0^{+2.7}_{-0.9}$ \\
\hline
$\rm [Fe/H] > 0$                                       & $137\,|\,30$        & 0.918      & $23.8^{+4.2}_{-3.4}$   &    $137\,|\,26$        & 0.951      & $20.0^{+3.9}_{-3.0}$ \\
\hline
\end{tabular}
\tablefoot{Same columns as Table~\ref{table_occurrence_rates_metstar}.}
\label{table_SP_Mixed_occurrence_rates}
\end{table*}

\clearpage
\newpage

~\\

\section{Occurrence rates of cold Jupiters in the AAT, CLS, and HARPS radial velocity surveys}

~\\

\renewcommand{\arraystretch}{1.3}

\begin{table*}[h]
\tiny
\centering
\caption{Occurrence rates of cold Jupiters in the CLS survey by using the selection criteria in BL24 (top) and this work (bottom).}   
\begin{tabular}{|c|c|c|c|c|c|c|}
\cline{2-7}
\multicolumn{1}{c|}{} & \multicolumn{3}{c|}{$\mp=0.3-13\,\Mjup$} & \multicolumn{3}{c|}{$\mp=0.5-20\,\Mjup$} \\ 
\hline
\multicolumn{1}{|c|}{[Fe/H]} & \multicolumn{1}{c|}{$N_\star\,|\,d$} & \multicolumn{1}{c|}{$C$} & \multicolumn{1}{c|}{$f_{\rm CJ}$ [\%]} &
\multicolumn{1}{c|}{$N_\star\,|\,d$} & \multicolumn{1}{c|}{$C$} & \multicolumn{1}{c|}{$f_{\rm CJ}$ [\%]} \\  
\hline
\hline
\multicolumn{7}{|c|}{\textbf{\normalsize CLS radial velocity sample (BL24)}} \\
\hline
\hline
$\overline{\rm [Fe/H]}=+0.008\pm0.004$   & $562\,|\,51$     & 0.981     & $9.2^{+1.9}_{-1.1}$     & $562\,|\,46$     & 0.993    & $8.2^{+1.3}_{-1.0}$ \\
\hline
\cline{1-7}
$\rm [Fe/H]<-0.1$                                       & $135\,|\,4$        & 0.980     & $3.0^{+2.3}_{-0.9}$     & $135\,|\,4$      & 0.993     & $3.0^{+2.3}_{-0.9}$ \\
\hline
$-0.1 \le \rm [Fe/H] \le +0.1$                      & $201\,|\,12$      & 0.985     & $6.1^{+2.2}_{-1.3}$     & $201\,|\,11$    & 0.994     & $5.5^{+2.1}_{-1.2}$ \\
\hline
$\rm [Fe/H]>+0.1$                                      & $226\,|\,35$      & 0.979     & $15.8^{+2.8}_{-2.1}$   & $226\,|\,31$     & 0.992     & $13.8^{+2.6}_{-2.0}$ \\
\hline
\hline
\multicolumn{7}{|c|}{\textbf{\normalsize CLS radial velocity sample (this work)}} \\
\hline
\hline
$\overline{\rm [Fe/H]}=+0.025\pm0.004$   & $402\,|\,50$       & 0.987      & $12.6^{+1.8}_{-1.5}$     & $402\,|\,45$     & 0.995     & $11.2^{+1.8}_{-1.4}$ \\
\hline
\cline{1-7}
$\rm [Fe/H]<-0.1$                                       & $95\,|\,4$           & 0.990      & $4.2^{+3.1}_{-1.2}$       & $95\,|\,4$         & 0.996     & $4.2^{+3.1}_{-1.2}$ \\
\hline
$-0.1 \le \rm [Fe/H] \le +0.1$                      & $149\,|\,12$       & 0.989      & $8.1^{+2.8}_{-1.7}$       & $149\,|\,11$       & 0.996    & $7.4^{+2.7}_{-1.6}$ \\
\hline
$\rm [Fe/H]>+0.1$                                      & $158\,|\,34$        & 0.985     & $21.8^{+3.6}_{-2.9}$     & $158\,|\,30$      & 0.994    & $19.1^{+3.5}_{-2.7}$ \\
\hline
\end{tabular}
\tablefoot{Same columns as Table~\ref{table_occurrence_rates_metstar}.}
\label{table_CLSsurvey_occurrence_rates}
\end{table*}

~\\
~\\

\renewcommand{\arraystretch}{1.3}

\begin{table*}[h]
\tiny
\centering
\caption{Occurrence rates of cold Jupiters in the AAT (top), CLS (middle), and HARPS (bottom) surveys.}   
\begin{tabular}{|c|c|c|c|c|c|c|}
\cline{2-7}
\multicolumn{1}{c|}{} & \multicolumn{3}{c|}{$\mp=0.3-13\,\Mjup$} & \multicolumn{3}{c|}{$\mp=0.5-20\,\Mjup$} \\ 
\hline
\multicolumn{1}{|c|}{[Fe/H]} & \multicolumn{1}{c|}{$N_\star\,|\,d$} & \multicolumn{1}{c|}{$C$} & \multicolumn{1}{c|}{$f_{\rm CJ}$ [\%]} &
\multicolumn{1}{c|}{$N_\star\,|\,d$} & \multicolumn{1}{c|}{$C$} & \multicolumn{1}{c|}{$f_{\rm CJ}$ [\%]} \\  
\hline
\hline
\multicolumn{7}{|c|}{\textbf{\normalsize AAT radial velocity survey}} \\
\hline
\hline
$\overline{\rm [Fe/H]}=-0.045\pm0.018$   & $196\,|\,20$     & 0.971     & $10.5^{+2.6}_{-1.8}$     & $196\,|\,20$     & 0.989     & $10.3^{+2.6}_{-1.8}$ \\
\hline
\cline{1-7}
$\rm [Fe/H]<-0.1$                                       & $69\,|\,3$         & 0.977     & $4.4^{+4.0}_{-1.4}$       & $69\,|\,3$         & 0.992      & $4.4^{+3.9}_{-1.3}$ \\
\hline
$-0.1 \le \rm [Fe/H] \le +0.1$                      & $66\,|\,5$         & 0.969     & $7.8^{+4.7}_{-2.2}$       & $66\,|\,5$         & 0.989     & $7.7^{+4.6}_{-2.1}$ \\
\hline
$\rm [Fe/H]>+0.1$                                      & $61\,|\,12$       & 0.966     & $20.4^{+6.2}_{-4.2}$     & $61\,|\,12$        & 0.988      & $19.9^{+6.1}_{-4.1}$ \\
\hline
\hline
\multicolumn{7}{|c|}{\textbf{\normalsize CLS radial velocity survey}} \\
\hline
\hline
$\overline{\rm [Fe/H]}=+0.025\pm0.004$   & $402\,|\,50$       & 0.987      & $12.6^{+1.8}_{-1.5}$     & $402\,|\,45$     & 0.995     & $11.2^{+1.8}_{-1.4}$ \\
\hline
\cline{1-7}
$\rm [Fe/H]<-0.1$                                       & $95\,|\,4$           & 0.990      & $4.2^{+3.1}_{-1.2}$       & $95\,|\,4$         & 0.996     & $4.2^{+3.1}_{-1.2}$ \\
\hline
$-0.1 \le \rm [Fe/H] \le +0.1$                      & $149\,|\,12$       & 0.989      & $8.1^{+2.8}_{-1.7}$       & $149\,|\,11$       & 0.996    & $7.4^{+2.7}_{-1.6}$ \\
\hline
$\rm [Fe/H]>+0.1$                                      & $158\,|\,34$        & 0.985     & $21.8^{+3.6}_{-2.9}$     & $158\,|\,30$      & 0.994    & $19.1^{+3.5}_{-2.7}$ \\
\hline
\hline
\multicolumn{7}{|c|}{\textbf{\normalsize HARPS radial velocity survey}} \\
\hline
\hline
$\overline{\rm [Fe/H]}=-0.122\pm0.011$    & $782\,|\,75$       & 0.968     & $9.9^{+1.2}_{-1.0}$     & $782\,|\,74$     & 0.984     & $9.6^{+1.2}_{-1.0}$ \\
\hline
\cline{1-7}
$\rm [Fe/H]<-0.1$                                       & $344\,|\,19$       & 0.970     & $5.7^{+1.5}_{-1.0}$       & $344\,|\,18$     & 0.986     & $5.3^{+1.5}_{-1.0}$ \\
\hline
$-0.1 \le \rm [Fe/H] \le +0.1$                      & $272\,|\,24$       & 0.963     & $9.2^{+2.1}_{-1.5}$       & $272\,|\,24$     & 0.981     & $9.0^{+2.1}_{-1.5}$ \\
\hline
$\rm [Fe/H]>+0.1$                                      & $166\,|\,32$       & 0.972     & $19.8^{+3.5}_{-2.8}$     & $166\,|\,32$     & 0.986     & $19.5^{+3.4}_{-2.7}$ \\
\hline
\end{tabular}
\tablefoot{Same columns as Table~\ref{table_occurrence_rates_metstar}.}
\label{table_RVsurveys_occurrence_rates}
\end{table*}

\twocolumn 

\end{appendix}

\end{document}